\documentclass[ reprint,superscriptaddress,amsmath,amssymb,aps,pre]{revtex4-2}

\usepackage{amsmath}
\usepackage{amssymb}
\usepackage{graphicx}
\usepackage{hyperref}
\usepackage[arrow, matrix, curve]{xy}
\usepackage{bm}
\usepackage{yhmath}
\usepackage{color}
\newcommand\diff{\mathrm{d}}

\usepackage[usenames,dvipsnames]{xcolor}
\hypersetup{colorlinks=true, linkcolor=BrickRed, urlcolor=blue!50!black, citecolor=blue!50!black}

\usepackage{amsmath,amsthm,amsfonts,amssymb,times,bbm,graphicx,color,epsfig,bm}
\usepackage{hyperref}

\usepackage[normalem]{ulem}

\newcommand\delete[1]{\textcolor{red}{  }}
\newcommand\revision[1]{  {#1}}

\makeatletter
\newcommand\hide@visible[1]{%
  \bgroup\fboxsep=.3ex\colorbox{Gray}{begin hide}%
  #1\colorbox{Gray}{end hide}\egroup%
}
\newcommand\hide@hidden[1]{%
  \bgroup\fboxsep=.3ex\colorbox{Gray}{hidden text}%
}
\newcommand\hide@invisible[1]{}
\newcommand\makevisible{\let\hide\hide@visible}
\newcommand\makehidden{\let\hide\hide@hidden}
\newcommand\makeinvisible{\let\hide\hide@invisible}

\makeatother
\makehidden

\begin{document}

{\tt published in J. Chem. Phys. 160, 224504 (2024)}
\title{Thermodynamic properties of quasi-one-dimensional fluids}


\author{Thomas Franosch${}^{*}$}
\affiliation{Institut f\"ur Theoretische Physik, Universit\"at Innsbruck, Technikerstra{\ss}e, 21A, A-6020 Innsbruck, Austria}
\email[corresponding author ]{thomas.franosch@uibk.ac.at}

\author{Rolf Schilling}
\affiliation{Institut f\"ur Physik, Johannes Gutenberg-Universit\"at Mainz,
 Staudinger Weg 7, 55099 Mainz, Germany}


\date{\today}

\begin{abstract}
We calculate  thermodynamic  and structural quantities of  a fluid of  hard spheres of diameter  $\sigma$ in a quasi-one-dimensional pore with accessible pore width $W $ smaller than $\sigma$ by applying  a perturbative method worked out earlier for a confined fluid  in a slit pore [Phys. Rev. Lett. \textbf{109}, 240601 (2012)]. 
In a first step, we prove that the  
 thermodynamic and a certain class of structural quantities of the hard-sphere fluid in the pore can be obtained from  a  purely
 one-dimensional fluid of rods of length $ \sigma$ with a central hard core of size  $\sigma_W =\sqrt{\sigma^2 - W^2}$ and a soft part  at both ends of length $(\sigma-\sigma_W)/2$. These rods  interact via effective $k$-body potentials $v^{(k)}_\text{eff}$ ($k \geq 2$) . The two- and the three-body potential will be calculated explicitly.  
 In a second step, the free energy of this effective one-dimensional  fluid  is calculated up to leading order in $ (W/\sigma)^2$. Explicit results for, e.g. the perpendicular pressure,   surface tension, and the density profile as a function of 
  density, temperature,  and pore width are presented presented and partly compared with results from Monte-Carlo simulations and standard virial expansions.  Despite the perturbative character of our approach it encompasses the singularity of the thermodynamic quantities at the jamming transition point. 
\end{abstract}

\maketitle


\section{Introduction}
Confined fluids play an important role in condensed matter physics, chemistry, and biology as well as in various technical applications. Therefore, they have attracted over decades the attention of numerous researchers studying, e.g., phase behavior and phase transitions (see, e.g., the 
reviews~\cite{Evans:Advances_28:1979, Evans:JCP_46:1990,Lowen:JoPCondMat_21:2009}),
structural \cite{Dietrich:PhysRep_260:1995} and dynamical properties (see, e.g., \cite{Klafter:Molecular_Dynamics:1989,Schilling:PRE_93:2016,Mandal:PRL_118:2017}), as well as glassy behavior~\cite{Lang:PRL_105:2010,Mandal:NatComm_5:2014}. 

Of particular interest are three-dimensional (3D) fluids of $N$ particles confined such that the fluids become quasi-one-dimensional (q1D). This occurs, e.g., for particles of diameter $\sigma$ in cylindrical pores or two-dimensional slit pores, both of length $L$ and accessible pore width $W$ smaller than $\sigma$. For point particles, $W$ coincides with the actual geometrical pore width. 
The condition $W < \sigma$ prevents particles to pass each other along the q1D pore. Strictly speaking, this is only true for hard particles. 

In contrast to 1D fluids, q1D fluids can be prepared experimentally (see, e.g., Refs.~\cite{Cui:JCP_116:2002, Chen:JACS_127:2005, Lin:JPChem_113:2009, Kofinger:PCCP_13:2011}). Their theoretical modeling requires accounting for the particle fluctuations perpendicular to the 1D direction. The constraint that particles can not pass each other simplifies strongly their theoretical modeling since the particle positions, e.g., in a cylindrical pore, can be labeled by the ordering along the 1D direction. 
This property, also called ``single-file'' condition, is identical to that of purely 1D fluids and 
implies that the total pair potential can be arranged such that it consists  of only interactions between neighboring particles.
The maximum number $M$ of interacting neighboring particles depends on the range of the pair potential, in particular, for hard spheres (hard disks) of diameter $\sigma$, only nearest-neighbor interactions ($M=1$) occur for $0 \leq W/\sigma < \sqrt{3}/2 =: W_1/\sigma$, while at most next-nearest neighbors interactions ($M=2$) are possible for $\sqrt{3}/2 \leq W/\sigma < 1 $. This property and the fact that colloidal liquids can be well modeled by hard particles makes their study particularly attractive. This is one of the reasons why q1D fluids of hard spheres (hard disks) have been investigated intensively.

Despite the q1D nature, an explicit closed form of thermodynamic quantities, such as the Helmholtz free energy, and equation of state, etc.\@ can not be obtained, in contrast to numerous purely 1D systems \cite{Mattis:Many_body_problem:1993,Takahashi:Thermodynamics:1999}.
However, besides computer simulations (see, e.g., Refs. \cite{Antonchenko:MolPhys_65:1988, Peterson:JCP_88:1988, Demi:MolSim_5:1991, * Demi:MolSim_7:1991, Mon:JCP_112:2000, Huerta:PhysRevRes_2:2020}), different analytical approaches exist requiring approximations or even numerical solutions. Without attempting completeness, we mention the transfer-matrix method (see e.g. Refs.~\cite{Barker:AustrJPhys_15:1962, Kofke:JCP_98:1993, Kamenetskiy:JCP_121:2004, Varga:JStatMech_11:2011, Godfrey:PRE_89:2014,Godfrey:PRE_91:2015,Yamchi:PRE_91:2015,Hu:MolPhys_116:2018,Hu:PhysRevRes_3:2021,
Montero:JCP_158:2023,Montero:JCP_159:2023}) 
as well as the exploitation of the convolution structure of the configurational part of the canonical partition function in Ref.~\cite{Gursey:MathProc_1:1950}. 
The latter method is only applicable to fluids confined in a 2D-slit
with periodic boundary conditions perpendicular to the confining
boundary~\cite{Wojciechowski:JCP_76:1982, Pergamenshchik:JCP_153:2020, Pergamenshchik:JML_387:2023}.
A more systematic approach is the use of standard viral expansions (see, e.g., Refs.~\cite{Glandt:AIChE_27:1981, 
Mukamel:JSMTE_2009:2009,Mon:PRE_97:2018, Mon:PhysicaA_556:2020, Montero:JCP_158:2023}), which, however, are only valid for low densities.  
Statistical geometry allows expressing the virial $p V$ by the averaged free volume and its averaged surface. 
In combination with the so-called cavity method, the pressure-density tensor
of q1D fluids was calculated, again for low density~\cite{Kim:JCP_134:2011}.
Last but not least, density-functional theory has frequently been applied not only to confined fluids in a slit-pore~\cite{Evans:Advances_28:1979,Lowen:JoPCondMat_21:2009}, but also to q1D fluids\cite{Percus:MolPhys_100:2002}.

Besides studying thermodynamic properties of q1D fluids of hard spheres/disks, their close packing (see, e.g., Refs.~\cite{Pickett:PRL_85:2000,Ashwin:PRL_102:2009}), glassy properties, and jamming have also been intensively investigated (see, e.g., Refs.~\cite{Bowles:PRE_73:2006,Ashwin:PRL_102:2009,Ashwin:PRL_110:2013,Godfrey:PRE_89:2014,
Robinson:PRE_93:2016,Zhang:PRE_102:2020,Zarif:PRE_104:2021}), emphasizing the importance of q1D systems of hard-core particles.

Except for the standard virial expansion, all of these analytical approaches mentioned above have not allowed  for deriving \emph{explicit} and \emph{exact} expressions, e.g., for the parallel pressure $p_\parallel(T,n,W)$ \revision{ or the density profile $\rho(|\mathbf{z} |)$} as a function of \delete{temperature $T$,} density $n=N/L$, and (accessible) pore width $W$. 
\revision{Although thermodynamic and  structural properties can be expressed exactly by the eigenvalues of the transfer matrix, these eigenvalues can not be obtained analytically as a function of $n$ and $W$. One has to discretize the transversal coordinates in order to obtain a finite-dimensional matrix. Then, in a second step, its diagonalization has to be done  numerically for \textit{each} value of $n$ and \textit{each} value of $W$. If the discretization is fine enough, the results from the transfer-matrix method will be rather precise. For a 2D slit pore and a cylindrical pore this was demonstrated in one of the earliest studies by comparing the numerical transfer-matrix solutions for five values of $W$ with the results from a Monte-Carlo simulation \cite{Kofke:JCP_98:1993}. }

 Our major motivation is to elaborate a systematic approach going significantly beyond the standard virial expansion and allowing us to calculate analytically thermodynamic and structural quantities as a function of $n$ and $W$. In a quite  natural way, our method will also encompass the divergence of various thermodynamic quantities at the closed-packing density $n_\text{cp}(W)$.

Since q1D fluids are close to 1D fluids, an obvious question arises whether it is possible to reduce the calculation of the thermodynamic and structural properties of the q1D fluid to that of a purely 1D fluid. 
\revision{The transfer-matrix method represents one possibility.  Eliminating the lateral degrees of freedom (d.o.f.) a 1D fluid is obtained for $W < W_1$ with effective interactions  $\sigma(\mathbf{z}_i,\mathbf{z}_{i+1}) = \sqrt{\sigma^2 - (\mathbf{z}_{i+1}-\mathbf{z}_{i})^2}$ between the transversal nearest-neighbor displacements $(\mathbf{z}_i,\mathbf{z}_{i+1})$ \cite{Kofke:JCP_98:1993}. Another possibility is to map the q1D fluid  to a Tonks gas of polydisperse hard rods by interpreting $\sigma(\mathbf{z}_i,\mathbf{z}_{i+1})$ as independent collision parameters with distribution function $\rho_c(\mathbf{z}_{i+1}-\mathbf{z}_{i})$ \cite{Post:PRA_45:1992}. However, as the authors stress themselves this mapping is not exact because correlations between the transversal d.o.f are neglected. }

For a 3D fluid comprised of hard spheres of diameter $\sigma$ in a slit pore, the present authors (together with S. Lang) have worked out an analytical framework which allows calculating systematically the explicit dependence of thermodynamic quantities on $T$, 2D density $N/A$ (with $A$ being the wall area) and accessible pore width $W < \sigma$~\cite{Franosch:PRL_109:2012, * Franosch:PRL_110:2013, 
* Franosch:PRL_128:2022} ($W$ was called $L$ in Ref.~\cite{Franosch:PRL_109:2012, * Franosch:PRL_110:2013,
* Franosch:PRL_128:2022}). This approach does not require the 2D density to be low. Its main observation is that with the decreasing accessible pore width, the coupling between the unconfined and confined degrees of freedom becomes weaker and weaker. Their coupling constant $(W/\sigma) ^2$ is identified as the smallness parameter. 
The unperturbed system is the purely 2D fluid of hard disks of diameter $\sigma$. This framework was also applied to calculate structural quantities, such as the $m$-particle density~\cite{Lang:JCP_140:2014}. A recent MC simulation~\cite{Jung:PRE_106:2022} has shown that the analytical result, e.g., for $m=1$, i.e., the density profile, is in  very good agreement with the MC results, even for moderate packing fractions and $W/\sigma = 0.75$, which is not much smaller than 
$(W/\sigma)_{\text{max}}=1$, the upper bound for the validity of our systematic analytical approach.

Based on the quasi-one-dimensionality, the present work will provide a physical picture, which allows us  to reduce exactly
the q1D fluid of hard spheres to a purely 1D fluid of rods with effective interactions.   This picture is the starting point  for the application of a  perturbative method~\cite{Franosch:PRL_109:2012, * Franosch:PRL_110:2013, 
* Franosch:PRL_128:2022} applied to a q1D fluid of hard spheres. The q1D fluid has the tremendous advantage that the thermodynamic and structural quantities such as the pair-distribution function of the 
unperturbed fluid, which is the Tonks gas, are known analytically. Since the validity of our perturbative  method does not require low densities it differs strongly from a standard virial expansion. It also differs completely from all analytical approaches developed so far for q1D fluids since it allows us to calculate the dependence on density and pore width explicitly.

The outline of our paper is as follows. In Sec.~\ref{Sec:Hard_Sphere_Fluid} 
 we first extend our earlier theoretical framework to a $d$-dimensional fluid of hyperspheres confined to $d_\perp$ directions. The case of a q1D fluid of hard particles is obtained for $d_\perp = d-1$. By exploiting the "single-file" condition, analytic progress is made for a q1D fluid in Sec.~\ref{Sec:quasi-one-dimensional} and we will present explicit expressions for the perpendicular and parallel pressure, the surface tension, and the density profile. The result for the parallel pressure will be compared with those from standard virial expansions and Monte-Carlo simulations.
A summary and conclusions follow in  Sec.~\ref{Sec:conclusions}. In order not to overload the main text with technical manipulations, several details are presented in appendices.

\section{Hard-sphere fluids in an isotropic confinement }\label{Sec:Hard_Sphere_Fluid}

In this section, we extend the theoretical framework worked out in Ref.~\cite{Franosch:PRL_109:2012, * Franosch:PRL_110:2013,
* Franosch:PRL_128:2022} to 
a $d$-dimensional fluid of $N$ hyperspheres of diameter $\sigma$ 
confined in $d_\perp$ perpendicular directions. This not only makes our presentation self-contained but also allows us to demonstrate its applicability for hard particles in a more general confinement and arbitrary dimensions. We restrict ourselves to the simplest case of a rotationally invariant confining boundary. In this set-up, there is only a single length scale characterizing
the confinement, which is the diameter $W$ of the accessible part of the pore. 
The diameter of the physical pore 
then equals $W+ \sigma$. 

The accessible region for the centers of the particles is then $V_\parallel \times V_\perp$, where $V_\parallel = [-L/2, L/2]^{d_\parallel} $ is a hypercube of dimension $d_\parallel := d-d_\perp$ centered at the origin and 
$V_\perp = \{ \mathbf{z}\in \mathbf{R}^{d_\perp}: |\mathbf{z}| \leq W/2\}$ is a $d_\perp$-dimensional hypersphere of radius $W/2$ also centered at the origin. For simplicity, we also denote  the volume of the hypercube by $V_\parallel$ and the volume of the hyperspherical confinement
by $V_\perp$. The thermodynamic limit $N\to \infty, L\to \infty$ with fixed density $n = N/L^{d_\parallel}$ is anticipated. 
Positions in the accessible regions naturally decompose into $\mathbf{r}= (\mathbf{x},\mathbf{z}) $ with a longitudinal (unconfined in the thermodynamic limit) component $\mathbf{x} \in V_\parallel$ and a transverse (confined) component $\mathbf{z}\in V_\perp$. We use the terms longitudinal, parallel, and unconfined degrees of freedom interchangeably, as well as the terms transverse, perpendicular, and confined. 
We shall also use the shorthand notation where the positions $\mathbf{r}_j =(\mathbf{x}_j, \mathbf{z}_j)$ are abbreviated by $j=(j_\parallel, j_\perp)$. The collection of all particle positions $(1\ldots N)$ will be referred to as the configuration. Similarly, the collection of parallel components $
(1_\parallel\ldots N_\parallel)$ constitutes the parallel configuration. 
The particles interact via a hard-sphere repulsion 
\begin{align}\label{eq:potential}
V(1\ldots N) &= \sum_{i<j} v_{\text{HS}}(ij; \sigma) ,
\end{align}
where the sum runs over all pairs of particles. The hard-core pair potential 
$v_{\text{HS}}(ij; \sigma) $ equals infinity for separations $|\mathbf{r}_i-\mathbf{r}_j| < \sigma$ and zero otherwise. The confinement is implicitly accounted for by restricting the perpendicular coordinates, $|\mathbf{z}_i| \leq W/2$ for $i=1,\ldots, N$. However, one can lift this constraint and introduce a smooth confining potential acting only on the perpendicular components, as well as, a smooth interparticle potential. The corresponding modifications are straightforward and are deferred to Appendix~\ref{Sec:Appendix_A}. In the following, we will restrict ourselves to the situation of a quasi-$d_\parallel$-dimensional fluid of hyperspheres, i.e., to accessible pore widths $0 \leq W < \sigma$.

The calculation of thermodynamic quantities consists of two steps. First, the perpendicular degrees of freedom (d.o.f.) are eliminated in order to derive an effective interaction potential of a fluid in $d_\parallel$ dimensions. In a second step the Helmholtz free energy of this effective $d_\parallel$-dimensional fluid will be calculated.

\subsection{Effective potential}\label{SubSec:effective_potential}
\begin{figure}
\includegraphics[angle=0,width=\linewidth]{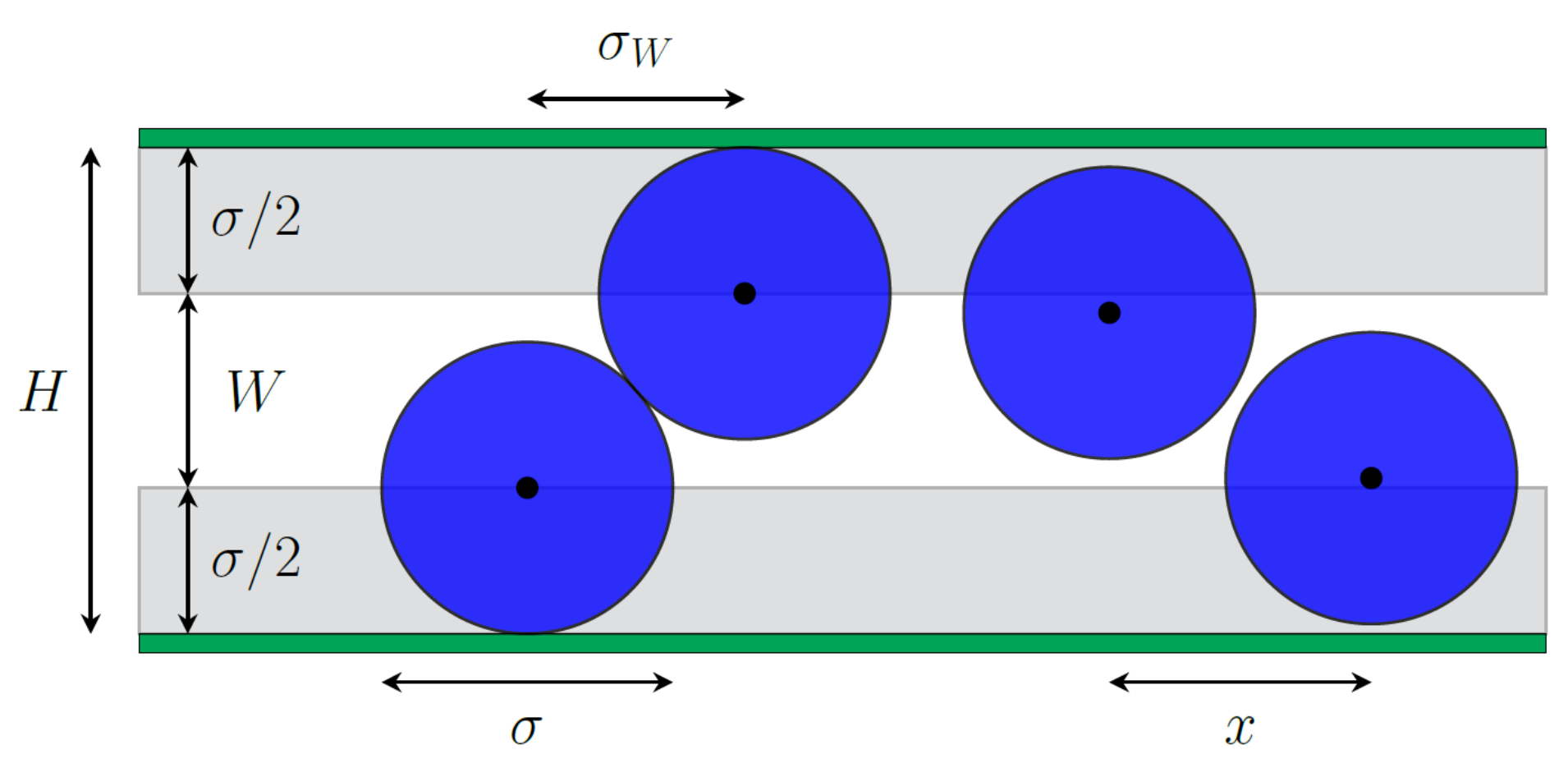}
\caption{Illustration of the q1D pore and two pairs of hard spheres with diameter $\sigma$. $H$ is the physical pore width, and $W= H - \sigma$ is the accessible one. Both hard spheres of the left pair are in contact with each other and with the confining boundary. Their centers have to lie in a cross section containing the pore axis, and their lateral distance $\sigma_W$ is the smallest possible lateral distance between two hard spheres. The pair on the right represents a general configuration with lateral distance $x$. Their centers do not have to lie in a plane containing the pore axis.
}\label{fig:q1D_cut}
\end{figure}

Consider an observable $\mathcal{O} = \mathcal{O}(1\ldots N)$ depending on the configuration $(1\ldots N)$. Then, we indicate the average over all confined components by angular brackets $\langle \ldots \rangle_\perp$,
\begin{align}
\langle \mathcal{O}(1\ldots N) \rangle_\perp = \left[ \prod_{j=1}^N \int \frac{\diff \mathbf{z}_j}{V_\perp} \right] \mathcal{O}(1\ldots N). 
\end{align}
Here, the integrals extend over the confinement volume $V_\perp$. 
In general, the average still depends on the parallel configuration $(1_\parallel\ldots N_\parallel)$. 
Considering a smooth one-particle potential $\sum_j U(\mathbf{z}_j)$ rather than a hard confinement is straightforward (see Appendix~\ref{Sec:Appendix_A}). 
We define the coarse-grained \revision{excess} free energy $\mathcal{F}(1_\parallel\ldots N_\parallel)$ depending only on the unconfined degrees of freedom $(1_\parallel\ldots N_\parallel)$
by averaging the Boltzmann factor over the confined directions (i.e., tracing out the perpendicular d.o.f.\@),
\begin{align}\label{eq:effpot1}
\exp[-\beta \mathcal{F}(1_\parallel\ldots N_\parallel ) ] 
&:= \langle \exp[- \beta V(1\ldots N) ] \rangle_\perp,
\end{align}
where $\beta = 1/k_B T $ denotes the inverse temperature. 
A crucial observation is that the smallest distance, $\sigma_W$, between two hard-sphere centers in the parallel direction is obtained when both spheres are in contact with each other and with the confining (rotationally invariant) boundary, which yields (see Fig.~\ref{fig:q1D_cut}).

\begin{figure}
\includegraphics[angle=0,width=\linewidth]{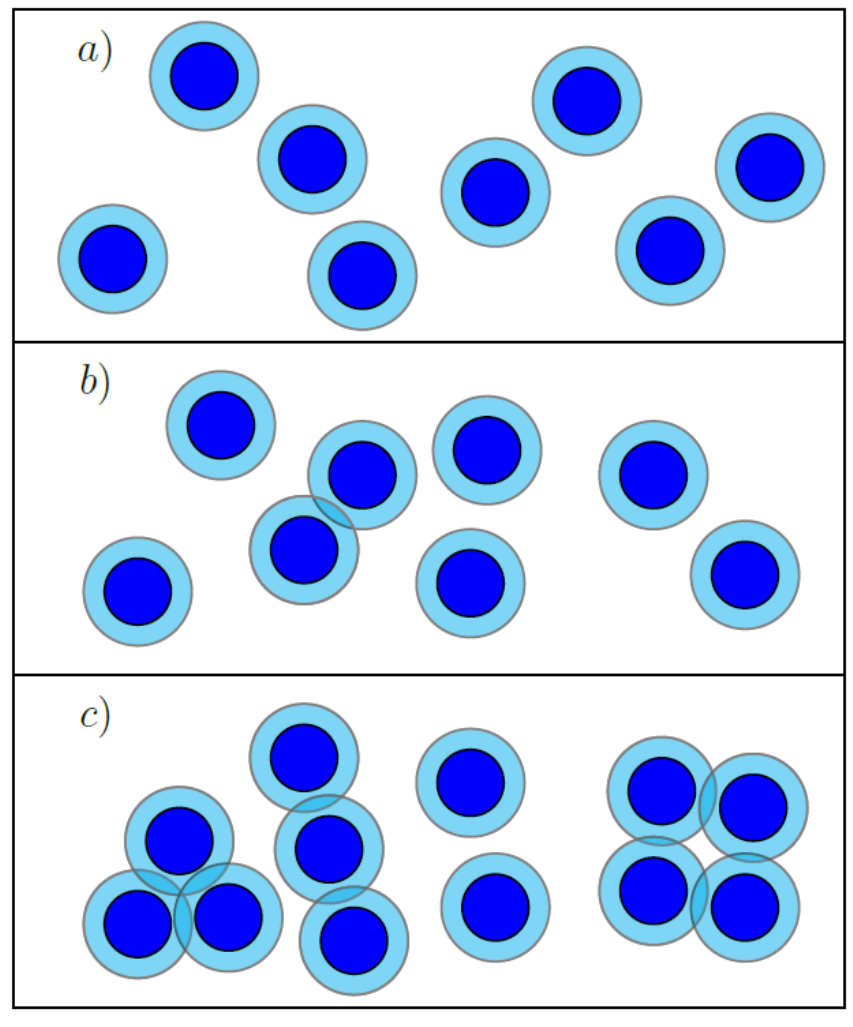}
\caption{Illustration of the cluster expansion for a quasi-two-dimensional fluid: The hard core with diameter $\sigma_{W}$ is shown in dark blue, and the light blue circle with diameter
$\sigma$ marks the range above which the excluded volume interactions vanish. (a) No clusters, (b) a two-cluster, (c) two three-clusters and a 
four-cluster immersed in the $d_\parallel$-dimensional fluid of spherical particles. }
\label{fig:cluster_expansion}
\end{figure}

\begin{figure}\includegraphics[angle=0,width=\linewidth]{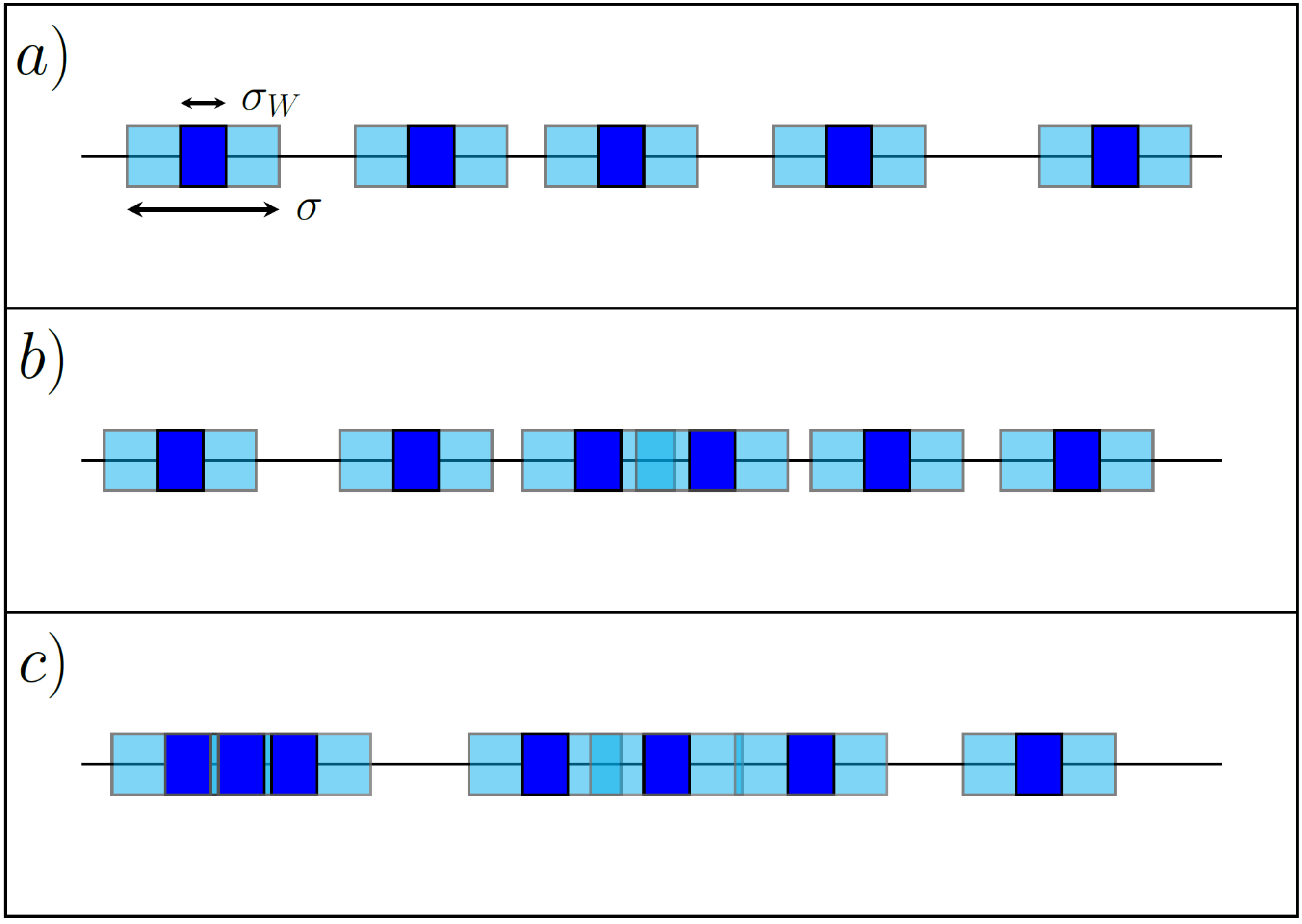}
\caption{Illustration of the cluster expansion for a q1D fluid. For better visibility, the rods of length $\sigma$ have a finite thickness.
The length $\sigma_{W}$ of their hard core is in dark blue. The light blue represents the soft part of each rod. (a) No clusters, (b) a two-cluster, (c) two three-clusters; note that for the left one, also the soft parts of both outer rods overlap.}
\label{fig:cluster_expansion-1d}
\end{figure}

\begin{align}\label{eq:width}
\sigma_W := 
\text{min} \,| \mathbf{x}_i-\mathbf{x}_j| = 
\sqrt{\sigma^2 - W^2} .
\end{align}
Therefore, we will choose the $d_\parallel$-dimensional fluid consisting of hard hyperspheres with diameter $\sigma_W$ as a \textit{reference fluid} and rewrite the Boltzmann factor in Eq.~\eqref{eq:effpot1} as follows:
\begin{equation}\label{eq:effpot2}
\exp[ - \beta V(1\ldots N) ] = 
\exp[-\beta V_0(1_\parallel\ldots N_\parallel) ] \prod_{i<j} [ 1 + f(ij)] ,
\end{equation}
with the hard-exclusion interaction with exclusion distance $\sigma_W$ in the parallel direction 
\begin{align}
V_0(1_\parallel\ldots N_\parallel) = \sum_{i<j} v_{\text{HS}}(i_\parallel j_\parallel; \sigma_W) ,
\end{align}
and the cluster function
\begin{align}\label{eq:cluster1}
f(ij) \equiv f(x_{ij}, z_{ij}) = \Theta( x_{ij}^2 + z_{ij}^2 - \sigma^2) 
- \Theta(x_{ij}^2 - \sigma_W^2).
\end{align}
Here, the Heaviside function, $\Theta(y)$, equals $0$ for $y < 0$ and $1$ for $y \geq0$. The cluster function is translationally invariant and only depends  on the distance $x_{ij} := |\mathbf{x}_i - \mathbf{x}_j |$ in the parallel 
and $z_{ij} := |\mathbf{z}_i-\mathbf{z}_i |$ in the perpendicular direction. Furthermore, $-1 \leq f(ij) \leq 0$ since $x_{ij}^2+ z_{ij}^2 > \sigma^2$ implies that $ x_{ij}^2 > \sigma_W^2$. 
Equation~\eqref{eq:effpot2} is an identity: its proof is straightforward using the fact that the Boltzmann factor for hard particles evaluates to zero or unity. Although Eq.~\eqref{eq:effpot2} resembles the standard virial expansion for low densities, it is quite different because in the latter the reference fluid is an ideal gas and not an interacting fluid. More importantly, the present cluster function is completely different since $f(ij) = 0$ for 
$x_{ij}> \sigma$ or $x_{ij} < \sigma_W$. Accordingly, the range of the parallel separation $x_{ij}$ where $f(ij)$ is nonzero is given by $\sigma - \sigma_W = O(W^2)$. In contrast, in the virial expansion, $f(ij)$ is nonzero for all separations smaller than the exclusion distance.
We define now as \emph{effective} potential the excess of the coarse-grained free energy with respect to the interaction energy of the
reference fluid. Making use of Eq.~\eqref{eq:effpot2}, 
we obtain from Eq.~\eqref{eq:effpot1}
\begin{align}\label{eq:effpot3}
V_\text{eff}(1_\parallel \ldots N_\parallel) & := \mathcal{F}(1_\parallel \ldots N_\parallel) - V_0(1_\parallel\ldots N_\parallel) \nonumber \\
&= - k_B T \ln \big\langle\prod_{i<j} [ 1 + f(ij) ] \big\rangle_{\perp}.
\end{align}
Since the support of $ f(ij)$ shrinks to zero for $W \to 0$, it can be considered as `small', in the sense of a measure. To simplify notation,
we abbreviate pairs $i<  j$ by $\alpha = (ij)$ and use a lexicographical order to define $\alpha< \beta$ occurring below. Then, with $y= \langle \prod_\alpha [1+ f(\alpha) ] \rangle_\perp -1$ and using $\ln (1+ y) = \sum_{k=1}^\infty (-1)^{k+1} y^k/k$, it follows that
\begin{align}\label{eq:effpot4}
\lefteqn{ V_\text{eff}(1_\parallel\ldots N_\parallel) = } \nonumber \\ 
=& - k_B T \sum_\alpha \left[\langle f(\alpha) \rangle_\perp - \frac{1}{2} \langle f(\alpha) \rangle_\perp^2 +\ldots \right] \nonumber\\
& - k_B T \sum_{\alpha < \beta } \left[\langle f(\alpha) f(\beta) \rangle_\perp - \langle f(\alpha) \rangle_\perp \langle f(\beta) \rangle_\perp +\ldots \right] \nonumber \\
&+\ldots \quad . 
\end{align}
The first line on the right-hand side consists 
of terms involving only single pairs $\alpha$. The second line collects contributions of two distinct pairs $\alpha< \beta$ and consists of products of averages of cluster functions involving the two pairs. Similarly, the $k$-th line consists of a sum over $k$ distinct pairs involving products of averages of cluster functions of the selected pairs. 
Except for the first line, all lines involve at least three particles. 
Already in the second line, many terms cancel, for example, if the pair $\alpha < \beta$ consists of four distinct particles the term explicitly displayed vanishes since $\langle f_\alpha f_\beta \rangle_\perp - \langle f_\alpha \rangle_\perp \langle f_\beta \rangle_\perp=0$. The argument transfers to arbitrary order such that in each line only those terms contribute where the pairs cannot be split into groups of distinct particles. 
These terms will be referred to as particle \emph{clusters} with non-vanishing contribution if each particle is separated by a distance no larger than $\sigma$ to at least one of the other particles of the cluster and no closer than $\sigma_W$ to any of the other particles of the cluster. In this sense, the clusters are connected. 
We can sort the terms into contributions involving clusters of an increasing number $k$ of particles; the effective potential is then a sum of 
effective $k$-particle interactions, $v^{(k)}_{\text{eff}}$, with a range of the maximal geometric extension of the cluster. The concept of $k$-clusters is illustrated schematically in Fig.~\ref{fig:cluster_expansion} for $d_\parallel = 2$ and for a q1D- fluid ($d_\parallel=1)$ in Fig.~\ref{fig:cluster_expansion-1d}.
Hence we have mapped the original confined $d$-dimensional fluid of $d$-dimensional hard spheres with diameter $\sigma$ to an unconfined $d_\parallel$-dimensional fluid of $d_\parallel$-dimensional spheres with a hard-core diameter $\sigma_W$ and a soft shell for separations between $\sigma_W$ and $\sigma$. While in the standard virial expansion further progress is made by classifying clusters into reducible and irreducible ones, no general progress appears to be possible in the current case.

The effective $k$-particle interaction contributing to $V_{\text{eff}}$ appears when the soft shells of $k$ adjacent $d_\parallel$-dimensional spheres overlap such that they form $k$-clusters (see Figs.~\ref{fig:cluster_expansion} and \ref{fig:cluster_expansion-1d}). We can make explicit progress by considering a configuration with only a single two-cluster consisting of particles 1 and 2 and all other particles separated by at least a distance $\sigma$ from any other. Then, Eqs.~\eqref{eq:effpot1}, \eqref{eq:effpot2}, \eqref{eq:effpot3} imply 
\begin{align}
\exp \left[ - \beta v^{(2)}_{\text{eff}}(1_\parallel 2_\parallel )\right] 
&= \left\langle \left[1 + f(12) \right] \right\rangle_\perp .
\end{align}
The effective two-particle potential is translationally and rotationally invariant in the parallel direction and only depends on the parallel distance $|\mathbf{x}_1-\mathbf{x}_2|$. 
Taking logarithms, we find 
\begin{align}\label{eq:effpot6}
v^{(2)}_\text{eff}(x) = - k_B T \ln \big[1+\langle f(x, z_{12}) \rangle_\perp \big] .
\end{align}
Here, the perpendicular average is taken over $\mathbf{z}_1, \mathbf{z}_2 $, and $x=|\mathbf{x}|$ denotes the parallel distance of the pair (see Fig.~\ref{fig:q1D_cut}). The preceding relation also  follows  directly by summing all terms in the first line of Eq.~\eqref{eq:effpot4} involving only a single pair. Since the cluster function is nonzero only in the small range of particle separations between $\sigma_W$ and $\sigma$, the second term in the square bracket of Eq.~\eqref{eq:effpot6} is non-vanishing only for these distances.

Similarly, considering a configuration consisting of a single three-cluster composed of particles 1, 2, and 3, and other particles in one-clusters only, a formal relation for the three-body potential 
$ v^{(3)}_{\text{eff}}(1_\parallel 2_\parallel 3_\parallel) $ can be derived,
\begin{align}\label{eq:eff-3-particle}
\exp & \left[ - \beta\left( v^{(2)}_{\text{eff}}(1_\parallel 2_\parallel) + v^{(2)}_{\text{eff}}(2_\parallel 3_\parallel) + v^{(2)}_{\text{eff}}(1_\parallel 3_\parallel) + v^{(3)}_{\text{eff}}(1_\parallel 2_\parallel 3_\parallel) \right) 
\right] \nonumber \\
= & \left\langle [ 1+ f(12) ] [1+ f(23) ] [1+ f(13)] \right\rangle_\perp .
\end{align}
In contrast to the well-known virial expansion, in general, $\langle f(12) f(13) \rangle_\perp \neq \langle f(12) \rangle_\perp \langle f(13) \rangle_\perp$, and therefore the three-particle contribution does not vanish even for configurations where two particles are apart at distances larger than $\sigma$. Therefore, in this sense, the clusters emerging in our expansion are already irreducible. 

\begin{figure}
\includegraphics[angle=0,width=\linewidth]{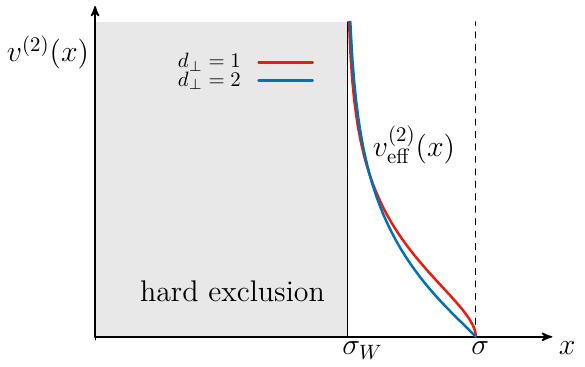}
\caption{ The total two-body potential $v^{(2)}(x) = v_{\text{HS}}(x) + v^{(2)}_{\text{eff}}(x)$ as a function of the lateral distance $x$ for the case $W/\sigma =0.75$. The gray region 
represents the excluded region corresponding to the closest lateral distance $\sigma_W$ two spheres can assume in confinement. The effective two-body potential shown for one ($d_\perp =1$) and two ($d_\perp =2$) confining directions occurs in the region $\sigma_W \leq x < \sigma$. \label{fig:soft_2_body_potential}}
\end{figure}

We have calculated the effective two-body potential for one ($d_\perp=1$) and two ($d_\perp=2$) confining directions (see Appendix~\ref{Sec:Appendix_B}). For $d_\perp=1$, it is given by Eq.~\eqref{eq:v2_Appendix},
\begin{align}\label{eq:effpot7}
v^{(2)}_\text{eff}(x) = - 2 k_B T \ln \big[1 - \sqrt{\big(\sigma^2 - x^2 \big)/W^2}\big] \ .
\end{align}
This case includes the situation of a 2D slit pore and coincides with our earlier result~\cite{Franosch:PRL_109:2012, * Franosch:PRL_110:2013,
* Franosch:PRL_128:2022}.
For $d_\perp=2$ the two-body potential $ v^{(2)}_\text{eff}(x)$ follows after substituting $1+\langle f(x, z_{12}) \rangle_\perp$
from Eq.~\eqref{eq:cluster_function_app} into Eq.~\eqref{eq:effpot6}. Since its expression is rather lengthy, it is not presented here. The total two-body potential $v^{(2)}(x) = v_\text{HS}(x) + v^{(2)}_\text{eff}(x)$, consisting of the hard-core exclusion $ v_\text{HS}(x)$ with diameter $\sigma_W$ and the effective potential $v^{(2)}_{\text{eff}}(x)$, is displayed in Fig.~\ref{fig:soft_2_body_potential} for both $d_\perp=1$ and $d_\perp=2$. It demonstrates that $v^{(2)}_{\text{eff}}(x)$ 
interpolates between the hard-core repulsion at distance $\sigma_W$ and the non-interacting regime $\geq \sigma$. This holds for any $d_\perp$. 
Since $\langle f(12) \rangle_\perp\to -1$ for $|\mathbf{x}_1-\mathbf{x}_2| \downarrow \sigma_W$
and $\langle f(12) \rangle_\perp = 0$ for $|\mathbf{x}_1-\mathbf{x}_2| \geq \sigma$, the second term in Eq.~\eqref{eq:effpot6} implies $v^{(2)}_{\text{eff}}(x) \to \infty$ for $x \downarrow \sigma_W$ and $v^{(2)}_{\text{eff}}(x) =0$ for $x \geq \sigma$. Note that the slope of $v^{(2)}_\text{eff}(x)$ at $x= \sigma$ diverges in the case of $d_\perp=1$, whereas it is finite for $d_\perp=2$. Probably, the latter holds for all $d_\perp \geq 2$.

\begin{figure}
\includegraphics[angle=0,width=1.0\linewidth]{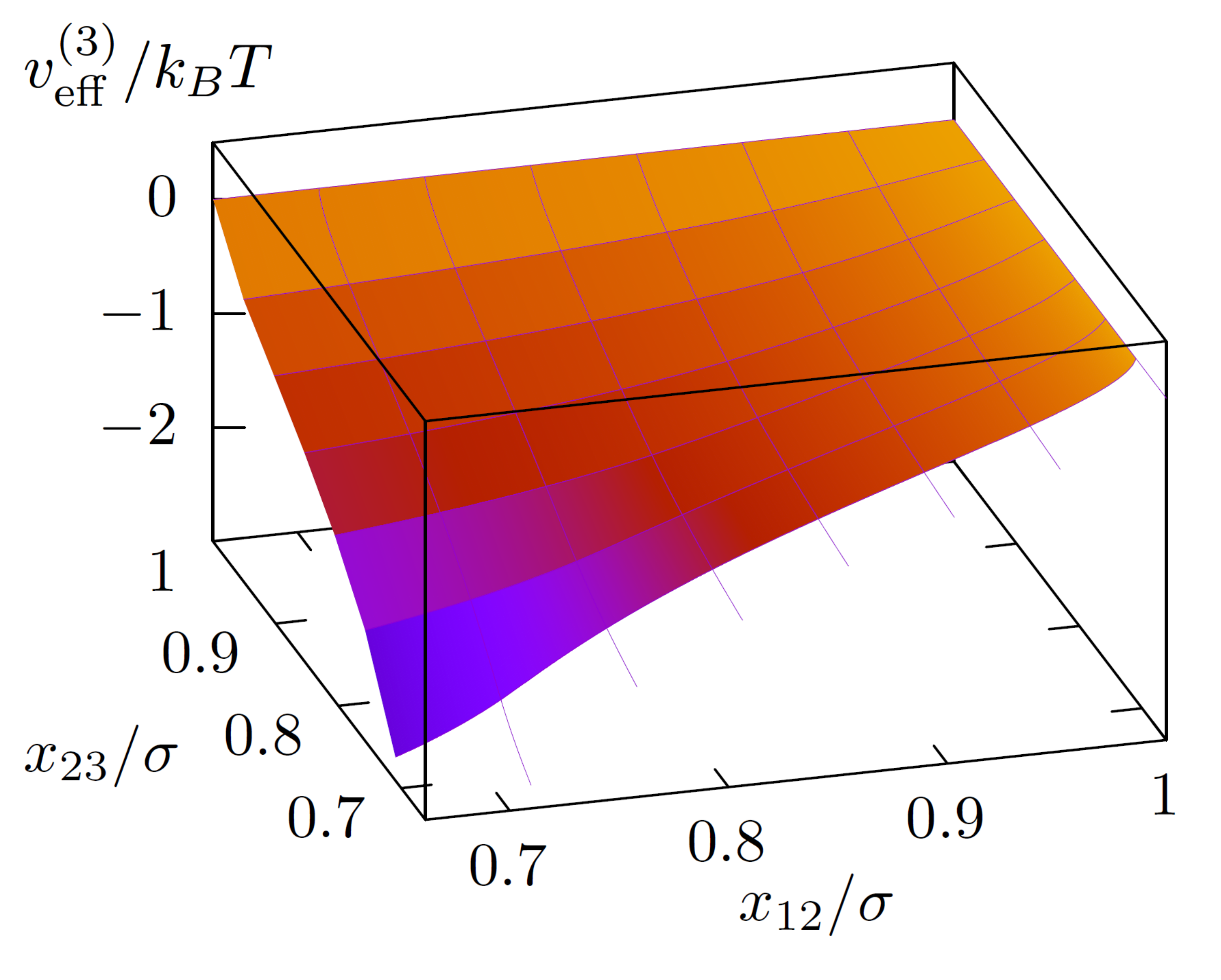}
\caption{ The effective three-body potential $v^{(3)}_{\text{eff}}(x_{12},x_{23})$ for 
$W/\sigma = 0.75$, i.e., $\sigma_W/\sigma \approx 0.66$, as a function of the lateral distances $x_{12}$ and 
$x_{23}$. \label{fig:soft_3_body_potential} }
\end{figure}

Equation~\eqref{eq:eff-3-particle} allows solving for $v^{(3)}_{\text{eff}}(\mathbf{x}_1 -\mathbf{x}_2,\mathbf{x}_2 -\mathbf{x}_3)$. Here, we restrict ourselves to the case for which the lateral distance between two of the hard spheres (spheres 1 and 3) is larger than $\sigma$, i.e., these spheres can never touch each other. This holds always for the q1D pore (our main concern) if $W < W_1$. Then, $v^{(3)}_{\text{eff}}(x_{12},x_{23})$ which only depends  on the lateral distances $x_{12}= | \mathbf{x}_1 -\mathbf{x}_2 |$ 
and $x_{23}= | \mathbf{x}_2 -\mathbf{x}_3|$ is 
calculated in Appendix~\ref{Sec:Appendix_C}.
From this analytical expression, it follows that $v^{(3)}_{\text{eff}}(x_{12},x_{23})$
converges to $- \infty$ for $x_{12} \to \sigma_W$ and $x_{23} \to \sigma_W$ and goes to zero for $x_{12} \to \sigma$ and/or $x_{23} \to \sigma$.
The graph of $v^{(3)}_{\text{eff}}(x_{12},x_{23})$ presented in Fig.~\ref{fig:soft_3_body_potential} is consistent with the properties mentioned above. Furthermore, it shows that the effective three-body potential is attractive for all $(x_{12},x_{23}) \in [\sigma_W, \sigma]^2$.
This implies that the 
sum of the repulsive two-body potentials overestimates the energy cost for forming a three-cluster. Clearly, the total energy for any $k$-cluster is positive.

\subsection{Free energy}\label{Sec:free_energy}

In a second step, we shall calculate the Helmholtz free energy of the confined $d$-dimensional fluid.
In contrast to bulk fluids the choice of the thermodynamic variables is not unique~\cite{Dong:CommPhys_6:2023}. There are three volumes: the total volume $V=V_\parallel V_\perp$, $V_\parallel$, and $V_\perp$. Here, we adopt the usual choice in computer simulations, i.e., we use $V_\parallel$ and $V_\perp$ as independent variables, although the latter is not necessarily macroscopic. 
By extensivity, the free energy is proportional to the particle number. Particle number $N$ and the parallel volume $V_\parallel$ enter in the free energy per particle $F/N$  only  via the  dimensionless number density $n^* := n \sigma^{d_\parallel} = N\sigma^{d_\parallel}/ V_\parallel$.  
Temperature $T$ does not play a significant role since the present fluid is an athermal system.
Then, the Helmholtz free energy $F= F(T,V_\parallel, V_\perp,N)$ can be represented as 
\begin{align}\label{eq:free-energy}
F(T,V_\parallel, V_\perp,N) = F^{\text{id}}(T,V,N) + F^{\text{ex}}(T,V_\parallel, V_\perp,N) ,
\end{align}
with the free energy of a $d$-dimensional ideal gas,
\begin{align}\label{eq:id-free-energy}
F^{\text{id}}(T,V,N) = -Nk_BT \ln\left[\frac{V}{N \lambda^d} \right],
\end{align} 
which involves the thermal wavelength $\lambda = \sqrt{2 \pi \hbar^2/ m k_B T}$.
The excess free energy $F^{\text{ex}}= -k_B T \ln Z^{\text{ex}} $
follows from the excess canonical partition function, 
\begin{align}\label{eq:ex-partition1}
Z^{\text{ex}} = Z^{\text{ex}}(T,V_\parallel, V_\perp,N)= \int \Big[ \prod_{j=1}^N \frac{\diff j}{V} \Big] \exp[- \beta V(1\ldots N) ].
\end{align}
Decomposing the Boltzmann factor into reference fluid and cluster functions [Eq.~\eqref{eq:effpot2}], we obtain
\begin{align}
Z^{\text{ex}} 
&= \int \Big[ \prod_{j=1}^N \frac{\diff j_\parallel}{V_\parallel} \Big] \exp[- \beta V_0(1_\parallel \ldots N_\parallel) ] \langle \prod_{i< j} [1+ f(ij) ] \rangle_\perp .
\end{align}
We define averages over the parallel degrees of freedom using the measure of the hard-sphere reference potential with exclusion distance $\sigma_W$,
\begin{align}\label{eq:definition_parallel_average}
\langle (\ldots) \rangle_\parallel := \frac{1}{Z^{\text{ex}}_\parallel } \int \Big[\prod_{j=1}^N \frac{\diff j_\parallel}{V_\parallel} \Big] \exp[- \beta V_0(1_\parallel\ldots N_\parallel)] ( \ldots ) ,
\end{align}
with the associated excess partition sum
\begin{align}
Z^{\text{ex}}_\parallel = Z^{\text{ex}}_\parallel(T, V_\parallel,N) 
= \int \Big[ \prod_{j=1}^N \frac{\diff j_\parallel}{V_\parallel} \Big] \exp[- \beta V_0(1_\parallel\ldots N_\parallel)].
\end{align}
Then, we find the exact relation 
\begin{align}\label{eq:ex-partition1}
Z^{\text{ex}} = Z^{\text{ex}}_\parallel \langle \langle \prod_{i<j} \big[1 + f(ij) \big] \rangle_\perp \rangle_\parallel,
\end{align}
or equivalently in terms of the free energies,
\begin{align}\label{eq:free_energy_decomposition}
F(T,V_\parallel,V_\perp,N) =& F^{\text{id}}(T,V,N) + F^{\text{ex}}_\parallel(T, V_\parallel, N; \sigma_W) \nonumber \\
&+ F^{\text{cluster}}(T,V_\parallel,V_\perp,N; \sigma_W) .
\end{align}
Here, $F^{\text{ex}}_\parallel = -k_B T \ln Z^{\text{ex}}_\parallel$ is the excess free energy of the $d_\parallel$-dimensional hard-sphere fluid for diameter $\sigma_W$ and 
\begin{align}\label{eq:cluster_free_energy}
F^{\text{cluster}} = - k_B T \ln \langle \langle \prod_{i<j} \big[1 + f(ij) \big] \rangle_\perp \rangle_\parallel,
\end{align}
accounts for the coupling of the perpendicular degrees of freedom to the parallel ones. 

The cluster free energy can be decomposed into contributions arising  
from cluster configurations with $(k-1)$  bonds
\begin{align}\label{eq:clusters_free_energy}
F^{\text{cluster}} = \sum_{k=2}^\infty F^{\text{cluster}}_k .
\end{align}
$F^{\text{cluster}}_2$ contains contributions from two-clusters, only. It is obtained
upon replacement $\langle \langle \prod_{i<j} \big[1 + f(ij) \big] \rangle_\perp \rangle_\parallel \mapsto \prod_{i<j} \big[1 + \langle \langle f(ij) \rangle_\perp \rangle_\parallel \big] $ in Eq.~\eqref{eq:cluster_free_energy}
\begin{align}\label{eq:ex-free-energy2}
F^{\text{cluster}}_2(T,V_\parallel, V_\perp,N) & = - k_B T\sum_{i<j} \ln\big[1 +\langle \langle f(ij) \rangle_\perp \rangle_\parallel \big] \nonumber\\
& = - k_B T \sum_{i<j} \langle \langle f(ij) \rangle_\perp \rangle_\parallel + (\cdots) \ ,
\end{align}
where $(\cdots)$ involves $ \sum_{i<j} [\langle \langle f(ij) \rangle_\perp \rangle_\parallel]^m, \ m \geq 2$. Their contribution vanishes in the thermodynamic limit.

The contributions of the higher clusters $F^{\text{cluster}}_k$ for $k > 2$
can be formally derived by treating the logarithm in Eq.~\eqref{eq:cluster_free_energy} similarly as described below Eq.~\eqref{eq:effpot3}. As an illustration, we present the result for $k=3$,  for k =3 obtained by exploiting the symmetry under permuting the particles
\begin{align}\label{eq:ex-free-energy3}
& F^{\text{cluster}}_3(T,V_\parallel, V_\perp,N) = -  3 k_B T \Big\{ \sum_{i<j<k} \big[\langle \langle f(ij)f(jk) \rangle_\perp \rangle_\parallel + \nonumber\\
&  - \langle \langle f(ij) \rangle_\perp \rangle_\parallel \langle \langle f(jk) \rangle_\perp \rangle_\parallel \big] + \nonumber\\
& + \sum_{i<j<k<l} \big[\langle \langle f(ij)f(kl) \rangle_\perp \rangle_\parallel - \langle \langle f(ij) \rangle_\perp \rangle_\parallel \langle \langle f(kl) \rangle_\perp \rangle_\parallel \big] + \nonumber\\
& + (\cdots) \Big\} \ ,
\end{align} 
where again $ (\cdots)$ involves terms such as $\sum_{i < j < k} \langle \langle f(ij) \rangle_\perp \rangle_\parallel \langle \langle f(jk) \rangle_\perp \rangle_\parallel \langle \langle f(ik) \rangle_\perp \rangle_\parallel $ which vanish in the thermodynamic limit. The term in the first square bracket arises from the three-clusters with two bonds [cf. Fig.~\ref{fig:cluster_expansion}(c)], and that in the second square bracket from a pair of two-clusters. Note that the expression in Eq.~\eqref{eq:ex-free-energy3}
do not have a counterpart in the standard virial expansion.

Since the support $[\sigma_W, \sigma]$ of the cluster functions shrinks to zero for vanishing pore width $W\to 0$, the cluster free energies fulfill $F^{\text{cluster}}_k = O\big((\sigma - \sigma_W)^{k-1}\big)$ leaving as dominant term the two-cluster contribution $F^{\text{cluster}}_2 = O(\sigma - \sigma_W) = O(W^2)$.
 With Eq.~\eqref{eq:ex-free-energy2}, we obtain the leading term of the cluster expansion,
\begin{align}\label{eq:clusterF}
F^\text{cluster}(T,V_\parallel, V_\perp,N) = & - k_BT\sum_{i<j} \langle \langle f(ij)\rangle_\perp \rangle_\parallel
\ + \nonumber\\
& + O(\sigma - \sigma_W)^2.
\end{align}

Now, there are two options to derive a perturbation series choosing  $W^2$ as the smallness parameter. 
If the reference fluid is the fluid of hard spheres with a diameter $\sigma_W$, as described above, all  coefficients $c_k(T,V_\parallel,N;\sigma_W)$ of $W^{2k}$ depend on $\sigma_W$, similar to the zero-order term, $ c_0(T,V_\parallel,N;\sigma_W) = F^{\text{ex}}_\parallel(T, V_\parallel, N; \sigma_W)$. We call this option
$\sigma_W$-expansion. However, if a hard-sphere fluid of diameter $\sigma$ (the diameter of the original confined $d$ dimensional fluid) is chosen,
 one has to  expand additionally $c_k(T,V_\parallel,N;\sigma_W)$ into a power series in $W^2$. This option shall be called $\sigma$-expansion.  From a purely mathematical point of view, the  $\sigma$-expansion is the systematic and consistent one since in the $\sigma_W$-expansion a variation of the smallness parameter $W$ also changes the coefficients  $c_k(T,V_\parallel,N;\sigma_W)$. However,  in  Sec.~\ref{Sec:quasi-one-dimensional}, it  will be shown why  the $\sigma_W$-expansion is closer to  physical reality. Nevertheless, in the following, we will choose the $\sigma$-expansion. Note, instead of  $W^2$, that one could also use $(\sigma - \sigma_W)$ as the smallness parameter. 
 
 $F^\text{cluster}$  can  be calculated  analytically in leading order in $W^2$ for all $d_\parallel \geq 1$  and all   $d_\perp \geq 1$ (see Appendix~\ref{Sec:Appendix_D})  with the result  
 \begin{align}\label{eq:clusterF-final} 
F^{\text{cluster}}  = &  N k_BT\frac{d_\perp+4}{d_\perp+2} \  \Omega_{\parallel}  n^*   g_+( n^*) 
\frac{W^2}{8\sigma^2}    +   O(W^4)    .
\end{align}
The geometrical prefactor involves $\Omega_\parallel= 2\pi^{d_\parallel/2}/\Gamma(d_\parallel/2)$, the surface area of a $d_\parallel$-dimensional hypersphere with unit-radius, where $\Gamma(\cdot)$ denotes the Gamma function. Furthermore, $F^{\text{cluster}}$ depends on density  via the product of $n^* $
 and the pair-distribution function, $g_+(n^*) := g(x\downarrow \sigma;n^*)$,  of the $d_\parallel$-dimensional
 hard-sphere fluid with diameter $\sigma$ at contact.  Note that  the contact value depends on the (dimensionless) density of the reference fluid and we shall  highlight this dependence throughout. 
For a  3D-fluid in a slit pore ($d_\parallel=2, d_\perp=1$)
 the result in Eq.~\eqref{eq:clusterF-final} coincides with our earlier result [cf. Eq.(8) in Ref.~\cite{Franosch:PRL_109:2012, * Franosch:PRL_110:2013,
* Franosch:PRL_128:2022}].

Since we have chosen the $\sigma$-expansion, we also have  to expand the free energy,
$ F_{\parallel}^\text{ex}(T,V_\parallel,N;\sigma_W)$, \delete{of the reference fluid} \revision{of the 
$d_\parallel$-dimensional fluid of hyperspheres with a diameter $\sigma_W$} to leading nontrivial order in 
$W$ which yields (see Appendix~\ref{Sec:Appendix_E}) 
\begin{align}\label{eq:parallelF-final2}
F_{\parallel}^{\text{ex}}(T,V_\parallel,N;\sigma_W) =& 
F_\parallel^{\text{ex}}(T,V_\parallel, N ; \sigma ) \nonumber \\
& - N k_B T \Omega_{\parallel} n^* g_+(n^* ) \frac{W^2}{4 \sigma^2} + O(W^4) .
\end{align}
Collecting results from Eqs.~\eqref{eq:free_energy_decomposition}, \eqref{eq:clusterF-final}, and \eqref{eq:parallelF-final2}, we find for the free energy
\begin{align}\label{eq:finalF1}
F(T,V_\parallel, V_\perp, N) =& F^{\text{id}}(T,V,N) + F^{\text{ex}}_\parallel(T,V_\parallel, N; \sigma) \nonumber \\
& + \sum_{k=1}^\infty F_k(T,V_\parallel, V_\perp, N) ,
\end{align} 
where the excess free energy in the first line corresponds to a $d_\parallel$-dimensional hard-sphere fluid with diameters $\sigma$. The second line collects the corrections due to the coupling of the unconstrained to the constrained d.o.f.\@ in ascending 
order of powers of the pore width: $F_k(T,V_\parallel, V_\perp, N) = O(W^{2k})$. The leading correction term $F_1= O(W^2)$ can be evaluated explicitly using the results of Eqs.~\eqref{eq:clusterF-final}, \eqref{eq:parallelF-final2}, 
\begin{align}\label{eq:finalF2}
\lefteqn{F_1(T,V_\parallel, V_\perp, N ) = } \nonumber \\
&= - N k_B T \Omega_{\parallel} \ n^* g_+(n^*) \frac{d_\perp}{8(d_\perp + 2)} \left( \frac{W}{\sigma} \right)^2. 
\end{align}
Note that the nontrivial density dependence of the pair-distribution function in $F_1(T,V_\parallel,N)$ demonstrates that the present approach differs completely from the standard virial expansion.

In Appendix~\ref{Sec:Appendix_C}, we address the case of a confined fluid of point particles with a smooth pair potential 
$v(ij)$ only depending  on the magnitude of the separation of a pair. 
Relying on a cumulant expansion~\cite{Hansen:Theory_of_Simple_Liquids:2023} we find that the total Helmholtz free energy is still provided by Eq.~\eqref{eq:finalF1} where $F_{\parallel}^{\text{ex}}(T,V_\parallel,N)$ is now the excess free energy of a
$d_\parallel$-dimensional reference fluid of point particles with pair potential $v( i_\parallel, j_\parallel)$ . The leading-order correction is worked out explicitly in Eq.~\eqref{eq:F-final3},
\begin{align}\label{eq:finalF1-point} 
\lefteqn{ F_1(T,V_\parallel, V_\perp, N) = } \nonumber \\
& = N \Omega_{\parallel} n \frac{d_\perp}{8 (d_\perp+2)} W^2 \int^{\infty}_{0} \diff x \, x^{d_\parallel -2} g(x) v'(x) .
\end{align}
Here, $g(x)$ is the pair-distribution function of the $d_\parallel$-dimensional reference fluid with smooth pair potential, 
Eq.~\eqref{eq:pair-distribution}. 
The leading correction approaches the result of Eq.~\eqref{eq:finalF2} if the pair potential approaches the hard-exclusion interaction. This can be elaborated similarly to the virial route for the pressure of a hard-sphere fluid
by introducing the cavity function~\cite{Hansen:Theory_of_Simple_Liquids:2023}.

\subsection{Density profile}\label{Sec:Density_profile}

The above theoretical framework can also be applied to calculate structural quantities, such as the $m$-particle density $\rho^{(m)}(1\ldots m)$. This was elaborated in Ref.~\cite{Lang:JCP_140:2014} for a 3D fluid in a slit pore. Restricting to the density profile, $\rho^{(1)}(1) $, we will extend this result to  
the $d$-dimensional fluid of hyperspheres in an isotropic confinement. In particular, we shall show that the density profile can be obtained much easier than in the approach in Ref. \cite{Lang:JCP_140:2014}. 

 The $m$-particle density  is a basic structural  entity characterizing the distribution of particles in a fluid. It is defined by~\cite{Hansen:Theory_of_Simple_Liquids:2023}
\begin{align}\label{eq:mpdensity}
\rho^{(m)}(1\ldots m) =\frac{N!}{(N-m)!} \int  \prod^N_{j=m+1} \diff j \,  \frac{ \exp[- \beta V(1\ldots N)]}{ V^N Z^{\text{ex}} }. 
\end{align}
Here, we restrict the discussion to the one-particle density. 
With the identity Eq.~\eqref{eq:effpot2}, we can rewrite the canonical averaging as 
\begin{align}\label{eq:profile1}
\rho^{(1)}(1) = \frac{N}{V} \frac{\langle \langle \prod_{i<j} [1+ f(ij) ] \rangle_\perp' \rangle_\parallel}{ \langle\langle   \prod_{i<j} [1+ f(ij) ] \rangle_\perp \rangle_\parallel} , 
\end{align} 
where  the prime  in the numerator indicates that there is no averaging over  particle 1 in the perpendicular direction.  Here, we used that 
by translational symmetry along the parallel direction, the one-particle density 
$\rho^{(1)}(1) \equiv \rho^{(1)}(\mathbf{x}_1,\mathbf{z}_1)$ does not depend on $\mathbf{x}_1$, so we can average over it $(\ldots ) \mapsto V_\parallel^{-1} \int \diff \mathbf{x}_1\, (\ldots) $.  Rotational symmetry in the perpendicular direction implies that it only depends  on the magnitude $z_1 = |\mathbf{z}_1|$ and we write   $\rho(z_1) := \rho^{(1)}(\mathbf{x}_1,\mathbf{z}_1)$.

As above, the numerator and denominator can be represented by a cluster expansion. Neglecting  $k$-clusters for $k\geq 3$ corresponds to substituting $\langle \langle \prod_{i < j} [1 + f(ij)]\rangle^{\alpha}_{\perp} \rangle_{\parallel}
 \mapsto    \prod_{i < j} [1 +\langle \langle f(ij)\rangle^{\alpha}_{\perp} \rangle_{\parallel} ]$ ($\alpha =$ prime or no prime). Since $\langle \langle f(ij)\rangle^{\alpha}_{\perp} \rangle_{\parallel} = O(W^2)$,
we can expand the product in Eq.~\eqref{eq:profile1} and obtain to leading order
\begin{equation}\label{eq:profile2}
\rho(z_1) = \frac{N}{V}  \big[1 + \sum^N_{j =2} \big(\langle \langle f_{1j}\rangle^{`}_{\perp} \rangle_{\parallel} - \langle \langle f_{1j}\rangle_{\perp} \rangle_{\parallel} \big) \big]  +O(W^4) .
\end{equation} 
 Substituting the result for $ \big(\langle \langle f(1j)\rangle^{`}_{\perp} \rangle_{\parallel} - \langle \langle f(1j)\rangle_{\perp} \rangle_{\parallel} \big) $ from Appendix \ref{Sec:Appendix_A}, it follows that 
\begin{align}\label{eq:profile_final}
\rho(z_1) =&   \frac{N}{V} \Big[ 1 +  \frac{n^* }{2} g_+(n^*)  \Omega_{\parallel} \sigma^{-2}   \left( z_1^2 -  \frac{d_\perp}{d_\perp+2} \frac{W^2}{4}\right)  \Big] \nonumber \\
& + O(W^4)  .
\end{align}
With the replacements  $W \mapsto L$ and the packing fraction $\varphi = \pi n \sigma^2/4$, this result is identical to our earlier result for a 3D hard-sphere fluid in a slit pore ($ d_\parallel =2, d_\perp =1$)  [cf. Eq.~(51) in Ref.~\cite{Lang:JCP_140:2014}].

Two further comments are in order. Equation~\eqref{eq:profile_final} implies first 
\begin{align}\label{eq:profile-av}
 \int \frac{\diff \mathbf{z}_1}{V_\perp} \rho(z_1) =  \frac{N}{V}   ,
\end{align}
as it should be, and second the validity of the contact theorem (see, e.g. \cite{Dong:CommPhys_6:2023}):
\begin{align}\label{eq:profile-contact}
k_B T \rho(W/2) =  p_{\perp} \equiv - \frac{1}{V_\parallel}\left. \partial F/ \partial V_\perp\right|_{T, V_\parallel, N}   \ .
\end{align}
The contact theorem  relates the perpendicular (transversal) pressure $p_\perp$ (see Sec.~\ref{Sec:quasi-one-dimensional}) to the density profile at contact, $z_1 = W/2$.

\section{Quasi-one-dimensional fluids}\label{Sec:quasi-one-dimensional}

In this section, we will apply the results of Sec.~\ref{Sec:Hard_Sphere_Fluid} to quasi-one-dimensional (q1D) fluids obtained by confining
a 3D fluid isotropically in $d_\perp=2 $ directions (cylindrical pore) or confining a 2D fluid in $d_\perp=1 $ directions (2D slit pore). The case of a q1D fluid has the considerable advantage that the excess free energy and the pair-distribution function for hard rods in one dimension (also called Tonks gas) are known exactly. Therefore we can identify the excess free energy of the one-dimensional reference fluid consisting of hard particles of diameter $\sigma$ with the well-known results of Refs.~\cite{Tonks:PhysRev_50:1936} and \cite{Salsburg:JCP_21:1953} 
\begin{align}\label{eq:tonks-F_ex}
F^{\text{ex}}_\parallel(T, L, N ; \sigma) = - N k_B T \ln( 1- n^*) ,
\end{align}
where we wrote $L$ rather than $V_\parallel$ for the one-dimensional case and the dimensionless density is $n^* = n \sigma= N \sigma/L$. The corresponding excess pressure for the Tonks gas then reads
\begin{align}\label{eq:ex_pressure_Tonks}
p^{\text{ex}}_\parallel(T,L,N ; \sigma) = n k_B T \frac{n^*}{1-n^*} .
\end{align}
The virial theorem~\cite{Hansen:Theory_of_Simple_Liquids:2023} relates the excess pressure directly to its contact value. Specializing Eq.~\eqref{eq:virial_theorem} to the one-dimensional case, we find
\begin{align}\label{eq:tonks-g}
g_+(n^*) =1/ (1-n^*) .
\end{align}

\subsection{Thermodynamic and structural quantities}

Substituting the results from Eqs.~\eqref{eq:tonks-F_ex} and \eqref{eq:tonks-g} into Eq.~\eqref{eq:finalF1} and using Eq.~\eqref{eq:finalF2} leads to the explicit expression 
\begin{align}\label{eq:series_F}
F(T,L,V_\perp, N) =& F^{\text{id}}(T,V,N) - N k_B T \ln(1-n^*) \nonumber \\ 
& - N k_B T \frac{n^*}{1-n^*} \frac{d_\perp}{4(d_\perp+2)} \left( \frac{W}{\sigma} \right)^2 \nonumber \\
&+ O(W^4),
\end{align}
where $V= L V_\perp = L \Omega_{\perp} (W/2)^{d_\perp}/d_\perp$ is the total volume of the pore with $\Omega_\perp$ denoting the surface of the $d_\perp$-dimensional unit sphere.

This is now an appropriate place coming back to both options to perform a perturbation 
series, discussed shortly in Subsection \ref{Sec:free_energy}. Equation~\eqref{eq:series_F} presents the perturbation expansion with $(W/\sigma)^2$ as the smallness parameter. The Tonks gas of hard rods of length $\sigma$ is the unperturbed system. 
Therefore, all thermodynamic quantities of the q1D hard-sphere fluid will inherit the singular behavior of that Tonks gas at its jamming transition point $n^*=n^*_\text{cp}=1$. This can be checked for all quantities calculated in the subsections below. 

Choosing the other option, i.e., to use the original reference fluid which is the Tonks gas with hard rods of diameter $\sigma_W$ as the unperturbed system and $(\sigma - \sigma_W)$ as the smallness parameter all thermodynamic quantities will inherit the singular behavior of the closed-packed Tonks gas at maximum density $n =n_\text{cp}(W)= 1/\sigma_W$, i.e. at $n^*= n^*_\text{cp}(W) =\sigma/\sigma_W \geq 1$, provided that $W \leq W_1$. 
The situation for $W_1 < W < \sigma$ is much more involved~\cite{Pickett:PRL_85:2000,Ashwin:PRL_102:2009} and will not be discussed here.

Since the exact expressions for the thermodynamic quantities of the q1D fluid of hard spheres of diameter $\sigma$ have to be singular at $n^*= n^*_\text{cp}(W)$, only the latter option is consistent with this general physical property.

The preceding result, Eq.~\eqref{eq:series_F}, for the free energy can now be used for the calculation of various thermodynamic quantities of interest.

\subsubsection{Parallel pressure}
\begin{figure}
\includegraphics[width=\linewidth]{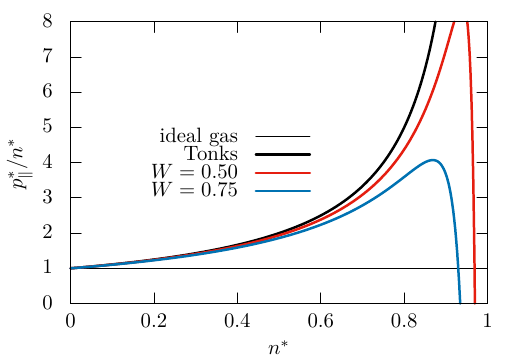}
\caption{Compressibility factor $p^*_\parallel/n^*= p_\parallel V/N k_B T$ in the q1D pore as a function of $n^*$. Shown are the results for a cylindrical pore, $d_\perp=2$, with dimensionless pore widths $W/\sigma = 0$ (Tonks gas), $0.50$, and $0.75$.
\label{fig:scaled_pressure}}
\end{figure}

We define the parallel pressure as the thermodynamic derivative with respect to the volume, while keeping the perpendicular volume $V_\perp$ fixed. Since $V= L V_\perp$, this entails
\begin{align}\label{eq:p-parallel}
\lefteqn{ p_\parallel(T,L,V_\perp, N) = - \frac{1}{V_\perp} \partial F/ \partial L|_{T,V_\perp,N} = } \nonumber\\ 
=& \frac{N k_B T}{V} \frac{1}{1-n^*} \left[1 - \frac{d_\perp}{4(d_\perp +2)} \frac{n^*}{1-n^*} \left(\frac{W}{\sigma} \right)^2 \right] + \nonumber \\
& + O(W^4) . 
\end{align} 
By construction, $p_\parallel V_\perp$ approaches the pressure of a Tonks gas, $n k_B T/(1-n^* )$, in the limit of $W\to 0$.
The compressibility factor $p_{\parallel} V/N k_B T$ for a cylindrical pore (i.e., $d_\perp = 2$) is displayed in 
Figure~\ref{fig:scaled_pressure} as a function of $n^*$ for various pore widths. 
In particular, one infers that for fixed $n^*$, it decreases with increasing pore width. 
The non-monotonic behavior for $W/\sigma= 0.75$, respectively, $0.50$ as a function of $n^*$ is most likely an artifact of truncating the expansion since the leading-order correction relative to the Tonks gas is no longer small, e.g., for $n^*=0.90$, it is $\approx 42\% $, respectively $18\%$.

The result for $p_{\parallel}$, Eq.~\eqref{eq:p-parallel}, reveals a divergence at the close-packing value
$n^*= n^*_\text{cp} =1$. Of course, our perturbative approach requires the second term in the square bracket of Eq.~\eqref{eq:p-parallel} to be much smaller than unity. Although for a fixed and small value of $W/\sigma$ this excludes $n^*$ to approach $n^*_\text{cp}$, our perturbative approach with the Tonks gas of hard rods of diameter $\sigma$ as the unperturbed system yields a singular behavior at $n^*_\text{cp} =1$. However, as discussed below in Eq.~\eqref{eq:series_F}, the physically correct result for $p_\parallel$ and for the other thermodynamic quantities below will be singular at $n^*= n^*_\text{cp}(W) > 1$.

\subsubsection{Perpendicular pressure}
We define the perpendicular pressure as the thermodynamic derivative with respect to the volume $V = L V_\perp$ while keeping the length $L$ fixed,
\begin{align}\label{eq:p-perp}
\lefteqn{ p_\perp(T,L,V_\perp,N) = - \frac{1}{L} \partial F/ \partial V_\perp |_{T,L,N} } \nonumber\\
= & \frac{N k_BT }{V} \left[1 + \frac{1}{2(d_\perp +2)} \frac{n^*}{1- n^*} 
\left (\frac{W}{\sigma} \right)^2 
\right] + O(W^4) ,
\end{align}
where in the second line we used that $ \partial / \partial {V_\perp} = (W/d_\perp V_\perp ) (\partial / \partial W) $. The leading-order term $ N k_BT/ V$ is the pressure of an ideal gas of $N$ particles in the volume
$V=L V_\perp $ of the q1D pore. 
The leading-order correction (second term in the square bracket) is always positive due to the repulsive nature of the hard core repulsion. 
In contrast to the corresponding term in Eq.~\eqref{eq:p-parallel} for $p_\parallel$, this term  sensitively
depends on the number of confining directions and vanishes for $d_\perp \to \infty$.

\subsubsection{surface tension}
The definition of the surface tension is somewhat subtle. An obvious way is to use temperature $T$, the pore volume $V = L V_\perp= L \Omega_{\perp} (W/2)^{d_\perp} /d_\perp$, the surface area of the confining boundary $\mathcal{A} = L \Omega_{\perp} (W/2)^{d_\perp-1}$, and the particle number $N$ as control parameters for the free energy and defining the surface tension as
\begin{align}\label{eq:tension-1}
\gamma = \left. \frac{\diff F }{\diff \mathcal A} \right|_{T,V,N} . 
\end{align}
\revision{Using the total derivative $(\diff F/\diff  \mathcal A)_V$ instead of a partial one indicates that $F$ is not a function of $ \mathcal A$ and $V$. }
This definition of $\gamma$ is often used in computer simulations (see, e.g. Ref.~\cite{Varnik:JCP_113:2000}). However, measuring this quantity \textit{in situ} seems not to be straightforward. In addition, the derivative of the free energy with respect to $\mathcal{A}$ at constant pore volume $V$ involves a variation of $F$
with the pore width $W$. For strong confinement the latter changes the density profile. This is also true if the corresponding grand potential is used. Therefore,
$\gamma$ as defined above is not an exclusively property of the surface. 

Using the definition in Eq. \eqref{eq:tension-1} requires expressing the free energy in terms of the variables $T,V, \mathcal{A}, N$ rather than $T, L, V_\perp, N$ as we used so far. Thus, we need to express the length and perpendicular volume in terms of the pore volume and surface. Elementary algebra yields
\begin{align}\label{eq:LandV} 
L &= L(V,\mathcal{A} ) = \frac{V d_\perp}{\Omega_{\perp}} \left( \frac{\mathcal{A}}{V d_\perp} \right)^{d_\perp} , \nonumber \\
V_\perp &=V_\perp(V, \mathcal{A} ) = \frac{\Omega_{\perp}}{d_\perp} \left( \frac{V d_\perp}{\mathcal{A}} \right)^{d_\perp} .
\end{align} 
Then by the chain rule
\begin{align}\label{eq:chain_rule}
\gamma =& (\partial F/\partial L)_{T, V_\perp, N} (\partial L/\partial \mathcal{A})_V \nonumber \\
&+ (\partial F/\partial V_\perp)_{T, L, N} (\partial V_\perp/\partial \mathcal{A} )_V .
\end{align}
The relations in Eq.~\eqref{eq:LandV} lead to the required partial derivatives
\begin{align}
(\partial L/\partial \mathcal{A} )_V &= d_\perp \frac{L}{\mathcal{A}} = \frac{W}{2 V_\perp}, \nonumber \\
(\partial V_\perp/\partial \mathcal{A} )_V &= - d_\perp \frac{V_\perp}{ \mathcal{A} } = - \frac{W}{2 L} , 
\end{align}
which implies for the surface tension the simple relation
\begin{align}\label{eq:gamma}
\gamma = \frac{1}{2} \big[- p_\parallel + p_\perp \big] W . 
\end{align}
Collecting results from Eqs.~\eqref{eq:gamma}, \eqref{eq:p-parallel}, and \eqref{eq:p-perp}, we find
\begin{align}\label{eq:tension-final}
\gamma =& - \frac{N k_B T}{V} \frac{W}{2} \frac{n^*}{1-n^*} \Big[ 1 -\frac{1}{4 (d_\perp +2)} \times\nonumber \\
&\times \left(2 + \frac{d_\perp}{1-n^*} \right) \left( \frac{W}{\sigma} \right)^2 + O(W^4) \Big].
\end{align}
Since $V_\perp \propto W^{d_\perp}$, $\gamma$ converges for the 2D-slit ($d_\perp =1$) to a constant in the limit $W \to 0$, while it diverges as $1/W$ for the cylindrical pore ($d_\perp =2$) due to the curvature of the confining walls. For $W$ and $n^*$ small enough, the surface tension as defined above is negative. This is a general property for  hard-sphere fluids~\cite{Heni:PRE_60:1999,Laird:JPhysChemC_43:2007}. This changes if the surface tension, denoted by $\gamma^{(p)}$,  is defined by the derivative of the free energy $F$ with respect to the physical surface area 
$\mathcal{A}^{(p)} = L \Omega_{\perp} [(W+\sigma)/2]^{d_\perp-1}$ at constant physical pore volume $V^{(p)}=L V^{(p)}_\perp = L (\Omega_{\perp} /d_\perp) [(W+\sigma)/2]^{d_\perp}$.
The superscript indicates the use of the real pore geometry. Then Eq.~\eqref{eq:LandV} is replaced by
\begin{align}\label{eq:LpandVp} 
L(V^{(p)},\mathcal{A}^{(p)} ) = \frac{V^{(p)} d_\perp}{\Omega_{\perp}} \left( \frac{\mathcal{A}^{(p)}}{V^{(p)} d_\perp} \right)^{d_\perp} , \nonumber \\
V_\perp(V^{(p)}, \mathcal{A}^{(p)} ) = \frac{\Omega_{\perp}}{d_\perp} \left( \frac{V^{(p)} d_\perp}{\mathcal{A}^{(p)}} - \frac{\sigma}{2} \right)^{d_\perp}    ,
\end{align} 
and Eq.~\eqref{eq:chain_rule}
 becomes
\begin{align}\label{eq:chain_rule-p}
\gamma^{(p)} =& (\partial F/\partial L)_{T, V_\perp, N} (\partial L/\partial \mathcal{A}^{(p)})_{V^{(p)}}
  \nonumber \\
&+ (\partial F/\partial V_\perp)_{T, L, N} (\partial V_\perp/\partial \mathcal{A}^{(p)} )_{V^{(p)}} .
\end{align} 
We will restrict ourselves to the physically relevant  2D slit pore ($d_\perp=1$) and  cylindrical pore ($d_\perp=2$). Then,
it follows that
\begin{align}\label{eq:gamma-p}
\gamma^{(p)} = \frac{1}{2} \alpha(W,d_\perp) \big[- p_\parallel W  + (W + \sigma ) p_\perp \big]    ,
\end{align}
with $\alpha(W,1) = 1$  and $\alpha(W,2)  = W/(W + \sigma)$.  For the slit pore, it follows  that
$\gamma^{(p)} - \gamma =  p_\perp \sigma/2$. The additional term $ p_\perp \sigma/2$ in $\gamma^{(p)}$
also occurs  for 3D fluids of hard spheres in a slit pore~\cite{Heni:PRE_60:1999, Laird:JPhysChemC_43:2007}. Substituting $ p_\parallel$ and $ p_\perp$ from Eqs.~\eqref{eq:p-parallel} and 
\eqref{eq:p-perp}, respectively, we obtain 
\begin{align}\label{eq:tension-final-p}
\gamma^{(p)} = & \  \frac{N k_B T}{2 V}   \alpha(W,d_\perp) \Big\{\big[ W+\sigma- \frac{W}{1-n^*} \big]  + \nonumber \\
& + \frac{n^*}{1-n^*} \big[ \frac{W+\sigma}{2 (d_\perp +2)}    + \frac{W d_\perp}{4 (d_\perp +2) } \frac{1}{1-n^*}\big ]  \left( \frac{W}{\sigma} \right)^2 + \nonumber \\
& + O(W^4)  \Big\}.
\end{align}
For $W/\sigma$ and $n^*$ small enough, the expression in the curly bracket is positive. Since its prefactor is positive, the surface tension $\gamma^{(p)}$  is also positive.  $\gamma^{(p)} > 0$ even holds for all 
$W/\sigma < 1$ if  $n^* <1/2 $, since the first square bracket in Eq.~\eqref{eq:tension-final-p} is positive. The second square bracket is always positive, provided that $n^* < 1$.

\subsubsection{density profile}
Last, we specialize the density profile, Eq.~\eqref{eq:profile_final}, to a q1D fluid relying on the known contact value, Eq.~\eqref{eq:tonks-g}, 
\begin{align}\label{eq:profile-q1d}
\rho(z_1) = n \left[ 1 + \frac{n^*}{1-n^*} \sigma^{-2} \left( z_1^2 - \frac{d_\perp}{d_\perp+2} \frac{W^2}{4} \right) 
\right] .
\end{align}
To leading order in $W$, the density profile is flat and acquires a parabolic shape by the leading-order correction.

\revision{\subsection{Comparison with results from other approaches}\label{Sec:Comparison} }

In this subsection we will compare our q1D results with results  from  Monte-Carlo simulations Ref. \cite{ Kofke:JCP_98:1993,Mon:PRE_97:2018}. This will be mainly illustrated for the parallel pressure of a cylindrical pore. Results for the perpendicular pressure and the surface tension seem not to exist. 
For $W/\sigma = 0.5$ we have taken  the data  from Fig. 3 of Ref.~\cite{ Kofke:JCP_98:1993} and   those for $W/\sigma = 0.8$ from Table II of Ref.~\cite{Mon:PRE_97:2018}. 
Since the Monte-Carlo results agree rather well with the numerically exact results from the transfer-matrix method, the comparison with the former will be sufficient. Furthermore, our result for the parallel pressure will also be compared with both virial expansions \cite{Mon:PRE_97:2018}.  The virial density  series for the dimensionless parallel pressure  $p_\parallel^* = p_\parallel V_\perp \sigma/k_B T$  reads 
\begin{align}\label{eq:n-series}
p_\parallel^*(n^*,W) = n^*\left[1 + \sum_{l=1}^{\infty} B_{l+1}(W) \  (n^*)^l  \right]   ,
\end{align}
and the virial pressure  series is given by 
\begin{align}\label{eq:p-series}
p_\parallel^*/n^* = 1 + \sum_{l=1}^{\infty} B'_{l+1}(W) \  (p_\parallel^*)^l      .
\end{align}
Here, the sequence of coefficients $( B_l')$ is uniquely determined by  $(B_l)$ by series inversion of Eq.~\eqref{eq:n-series} to $n^* = n^*( p_\parallel^*, W)$, and then expanding  $1/n^*$ in powers of  $p_\parallel^*$~\cite{Dobbins:JPhysChemRef_17:1988}.  For instance, $B'_2(W) = B_2(W)$ and   $B'_3(W) = B_3(W) - B_2(W)^2$. The coefficients  for $l \leq 3$ have been calculated analytically up to order $O(W^m)$ for $m \leq m_{\text{max}}(l)$ where  $ m_{\text{max}}(2)= 34$ and $ m_{\text{max}}(3)= 38$  \cite{Mon:PRE_97:2018}. 
The calculation of $B_4(R)$ and $B'_4(R)$ at which the series were truncated required a numerical computation (Ref.~\cite{Mon:PRE_97:2018}). 

Using Eq.~\eqref{eq:p-parallel}  it is an  easy exercise to  expand 
$p_\parallel^*(n^*,W)$ with respect to $n^*$. Comparing with the virial density series we obtain the exact result
\begin{align}\label{eq:n-series-W2}
B_l(W) = 1 - \frac{1}{8}(l-1)(W/\sigma)^2  + O(W^4) ,
\end{align}
for all $l$.  For $l = 2,3$, this result coincides with the analytical result  in Ref.~\cite{Mon:PRE_97:2018} up to 
 $O(W^2)$ [note that in Ref.~\cite{Mon:PRE_97:2018} $\hat{R} = W/2 \sigma$].
Using the known relations~\cite{Dobbins:JPhysChemRef_17:1988} to convert  the $(B_l)$ series into the $(B_l')$ series,  we find
\begin{align}\label{eq:p-series-W2}
B'_2(W) = 1 -\frac{1}{8}(W/\sigma)^2    \ ,  \  B'_l(W) = O(W^4)  , \ l \geq 3  \  .
\end{align}
The result for $B'_2(W)$ is obvious because  $B'_2(W) \equiv B_2(W)$,  and the result for $B'_3(W)$
is consistent with its series  expansion  in Ref.~\cite{Mon:PRE_97:2018} since its  leading-order term is  
$(W/\sigma)^4$. We conjecture that $B'_l(W) = O(W^{2l-2})$.

The fact that already the leading-order result, Eq.~\eqref{eq:p-parallel}, of our systematic expansion allows determining \textit{all} coefficients of the virial density expansion up to order $O(W^2)$ proves that our result corresponds to summing up  an infinite number of terms of  the virial density series.  The corresponding sum  produces the denominators $(1- n^*)$ in  Eq.~\eqref{eq:p-parallel}. This underlines  the strength of our method and confirms that it is not a low-density expansion.
\begin{figure}
\includegraphics[angle=0,width=1.0\linewidth]{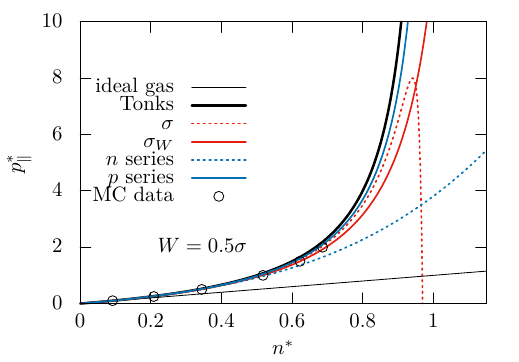}
\caption{ Dimensionless parallel pressure $p_\parallel^* = p_\parallel V_\perp \sigma/ k_BT$ for a cylindrical pore with accessible width $W= 0.5 \sigma$ as a function of the dimensionless density $n^* = n\sigma$. Besides the ideal gas and the Tonks gas, the results from our $\sigma$-expansion, $\sigma_W$-expansion, and both virial series \cite{Mon:PRE_97:2018} are displayed, as well as the Monte-Carlo data \cite{Kofke:JCP_98:1993} (open circles).\label{fig:comparison_Kofke}  
}
\end{figure}
\begin{figure}
\includegraphics[angle=0,width=1.0\linewidth]{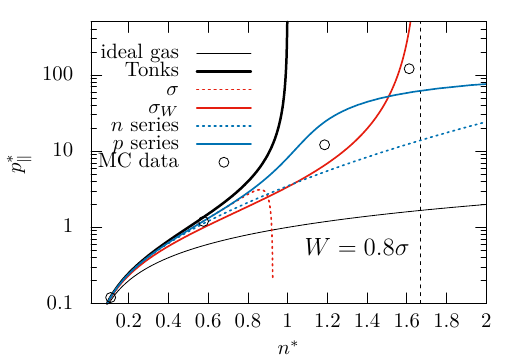}
\caption{Dimensionless parallel pressure $p_\parallel^* = p_\parallel V_\perp \sigma/k_BT$ for a cylindrical pore with accessible width $W= 0.8 \sigma$ as a function of the dimensionless density $n^* = n \sigma$ on a log-lin scale.  Besides  the Tonks gas, the results from our $\sigma$-expansion, $\sigma_W$-expansion, and both virial series \cite{Mon:PRE_97:2018} are displayed, as well as, the Monte-Carlo data \cite{Mon:PRE_97:2018} (open circles). The vertical dashed line marks the closed packing value
$n^*_\text{cp}(W) = \sigma/\sigma_W$ for $W/\sigma = 0.8$. 
 \label{fig:comparison_Mon} }
\end{figure}

Now we  compare our result, Eq.~\eqref{eq:p-parallel}, for $p_\parallel$ from the  $\sigma$-expansion with the results from both truncated virial series~\cite{Mon:PRE_97:2018} and the Monte-Carlo simulation~\cite{ Kofke:JCP_98:1993,Mon:PRE_97:2018}. The comparisons for a cylindrical pore ($d_\perp= 2$) with pore width $W= 0.5 \sigma$ and $W= 0.8 \sigma$ are shown in Fig.~\ref{fig:comparison_Kofke} and Fig.~\ref{fig:comparison_Mon}, respectively.
Figures~\ref{fig:comparison_Kofke} and \ref{fig:comparison_Mon} also include the result 
\begin{align}\label{eq:p-parallel_W}
\lefteqn{ p_\parallel(T,L,V_\perp, N) = - \frac{1}{V_\perp} \partial F/ \partial L|_{T,V_\perp,N} = } \nonumber\\ 
=& \frac{N k_B T}{V} \frac{1}{1-  n^*_W} \left[1 + \frac{3}{8} \frac{  n^*_W}{1-  n^*_W} \left(\frac{W}{\sigma} \right)^2 \right] + \nonumber \\
& + O(W^4) . 
\end{align} 
for the cylindrical pore ($d_\perp= 2$) from our   $\sigma_W$-expansion, which holds  for $W < W_1 = \sqrt{3}/2 \ \sigma $.  Here, we used the shorthand notation $n^*_W := n \sigma_W =  (\sigma_W/\sigma)n^*$. This result follows by  using the excess free energy $F^{\text{ex}}_\parallel(T, L, N ; \sigma_W) = - N k_B T \ln[ 1-  n^*_W] $ for the Tonks gas of hard rods of length $\sigma_W$ and the two-cluster result in leading order in $W^2$ (see  
Appendix~\ref{Sec:Appendix_D})
\begin{align}\label{eq:F_2-cluster_W}
F_1(T,L,V_\perp, N; \sigma_W) = & N k_B T  \frac{3}{8} \frac{  n^*_W }{1- n^*_W} \left(\frac{W}{\sigma} \right)^2  \ .
\end{align}

Figure~\ref{fig:comparison_Kofke} demonstrates that, except of the virial density expansion,  all other  expansions 
agree satisfactorily with the Monte-Carlo data, up to its maximum density $n^*_\text{max} \approx 0.69$. Although  the effective smallness parameter $[n^*/(1- n^*)] (W/\sigma)^2$ of the $\sigma$-expansion becomes $0.56$ for  $n^*_\text{max}$ and  is not very small, the leading-order result of the  $\sigma$-expansion is superior to the virial density expansion where terms even up to the fourth order have been taken into account.

The Monte-Carlo simulation in Ref.~\cite{Mon:PRE_97:2018} was performed at constant pressure which was increased exponentially. Therefore, the parallel pressure  in Fig. ~\ref{fig:comparison_Mon} is displayed on  a log-scale. Up to the density $n^* \approx 0.57$ of the Monte-Carlo data, again all expansions agree more or less well with the Monte-Carlo result. Of course, due to the  logarithmic scale, the deviations between the various results appear smaller.
The result from the truncated  virial pressure series  and the $\sigma$-expansion appears almost identical up to $n^* \approx 0.8$.  This value is a bit below the density where  the graph for the  $\sigma$-expansion
starts to decrease, indicating the importance of higher-order terms.

However, our main motivation to present  Fig.~\ref{fig:comparison_Mon}  is  not to check the quantitative accuracy of the various methods, but its qualitative behavior, particularly for higher densities up to the maximum closed-packing value
$n^*_\text{cp}(W) = \sigma/ \sigma_W$, which is valid for $W < W_1 $.  For $W/\sigma = 0.8$ (which is below $W_1/\sigma \approx 0.866$), it follows that $n^*_\text{cp}(W)\approx 1.66$ which is the position of the vertical dashed line in    Fig.~\ref{fig:comparison_Mon}.  At the Monte-Carlo points corresponding to the two largest densities,  $n^* \approx 1.18$  and $n^* \approx 1.61$,  the result from the truncated virial pressure  series  and the first-order result of the $\sigma_W$-expansion are the closest ones. However, approaching   $n^*_\text{cp}(W) \approx1.66$  the truncated virial pressure series  fails to describe the divergence at the jamming density $n^*_\text{cp}(W)$.  Truncating the virial pressure series, Eq.~\eqref{eq:p-series}, above $l=1$, one can prove that the parallel pressure converges for $n^* \to \infty$ to a constant due to the negative sign of $B'_l$ for $l=3, 4$~\cite{Mon:PRE_97:2018}.
Although  the density can not exceed $n^*_\text{cp}\approx 1.66$, this finding proves that the truncated virial pressure series fails to display the divergence of the parallel pressure at  $n^*_\text{cp}$ in striking contrast to the leading-order result of the  $\sigma_W$-expansion. This is not an artifact of the first order  but holds for all orders in $(W/\sigma)$ since all orders inherit this divergence from the Tonks gas with hard rods of length $\sigma_W$ as a reference fluid. A posteriori,  this justifies the relevance of the discussion  below 
Eq.~\eqref{eq:clusterF}.

The authors of Ref.~\cite{ Kofke:JCP_98:1993}  also presented the density profile for various values of the pressure and of $W/\sigma$.
Particularly for smaller values  of $W/\sigma$, their original figures are too small in order to deduce the Monte-Carlo points.  Nevertheless, we can check consistency with our result, Eq.~ \eqref{eq:profile-q1d}, which up to first order in $(W/\sigma)^2$ predicts a parabolic profile becoming more and more flat with decreasing pore width. 
Since the density profile for a cylindrical pore  in Figure 4 of  Ref.~\cite{ Kofke:JCP_98:1993} is plotted vs  $(z_1)^2$, a parabolic profile becomes  a straight line with a slope decreasing with decreasing $W/\sigma$.
For the two smallest pore widths, $W/\sigma = 0.25$  and $W/\sigma = 0.50$, the profile  indeed is a straight line 
for all pressures, with small deviations at the highest pressure (i.e., highest density). For $W/\sigma = 0.75$  and
$W/\sigma =1.0 $ (where our expansion loses its validity), the profile is still a straight line, at least for the smallest density.

In order to investigate the quality and validity of our results more systematically, further Monte-Carlo simulations are  necessary. This would also allow one  to perform a quantitative comparison  for the density profile, as done for a 3D fluid of hard  spheres in a slit pore \cite{Jung:PRE_106:2022}.

  \section{Summary and conclusions}\label{Sec:conclusions}

It appears obvious that the calculation of thermodynamic  quantities of  quasi-one-dimensional (q1D) fluids can be reduced to that of a purely one-dimensional (1D) fluid.
One of our main motivations has been to perform this mapping explicitly for a fluid of hard spheres of diameter $\sigma$ in a pore of physical width $H$. 
For accessible pore widths $W= H - \sigma  < \sigma$, we have proven that the thermodynamic and  certain structural properties of the q1D fluid can be obtained from a 1D fluid of rods of length $\sigma$ with a central hard core of size $\sigma_W = \sqrt{\sigma^2 - W^2 }$ and a soft part at both ends of length $(\sigma -\sigma_W )/2$ (see Fig.~\ref{fig:cluster_expansion-1d}). These rods interact via effective $k$-body potentials $v^{(k)}_\text{eff}(x_1,\ldots,x_N)$ obtained by eliminating the transversal (the confined ones) degrees of freedom (d.o.f).   $v^{(2)}_\text{eff}(x)$ and $v^{(3)}_\text{eff}(x_{12},x_{23})$ were calculated explicitly. 
The corresponding forces, i.e., the gradient of the effective potentials, are of an entropic  origin.
This is quite similar to the well-known depletion forces of particles in a solution of macromolecules, the Asakura-Oosawa model,  where the d.o.f\@ of the macromolecules are eliminated~\cite{Asakura:JCP_22:1954,Asakura:JPS_33:1958}. In contrast to our q1D fluid the two-body potential in the Asakura-Oosawa model is attractive. 
We stress that our exact mapping differs strongly not only from the approximate one to a Tonks gas with polydispersity~\cite{Post:PRA_45:1992}, but also from the exact mapping by eliminating the lateral d.o.f., which leads to the transfer-matrix approach for q1D fluids~\cite{ Kofke:JCP_98:1993}.

 The other major goal was to use our mapping to a 1D fluid   for the   calculation of  thermodynamic quantities of the q1D fluid, e.g., the Helmholtz free energy $F$.  This was achieved by a cluster expansion for hard spheres (or a cumulant expansion for  particles with smooth pair potentials)   with $(W/\sigma)^2$  as the smallness parameter.

Two options exist. Either the Tonks gas with hard rods of length $\sigma_W$ or $\sigma$ is used as the reference fluid. This leads to what we called $\sigma_W$- and $\sigma$-expansion,
 respectively. Its $k$-th term stems from all cluster configurations (see Fig.~\ref{fig:cluster_expansion-1d}) with $(k-1)$  bonds. 
 Its calculation involves the $k$-particle distribution function $g^{(k)}(x_{12},...,x_{k-1,k})$.
 Due to the one-dimensionality of our reference fluid and the nearest-neighbor interactions (for $W < W_1 = \sqrt{3} \sigma/2$) it factorizes into $\prod^{k-1}_{i=1 } g(x_{i,i+1})$ \cite{Salsburg:JCP_21:1953} where $g(x)\equiv g^{(2)}(x)$. Consequently, the $(k-1)$  bonds contribute  
 to $ \langle \prod^{k-1}_{i=1 } \big[n \int \diff x  \,  g(x )  f( x, z_{i,i+1}) \big] \rangle_\perp $. 
 Similarly to $k=2$ (which is discussed in  Appendix~\ref{Sec:Appendix_D}) one can show that 
 the $k$-th term of the $\sigma$-expansion will be proportional to $[\big(n^*/(1- n^*)\big) (W/\sigma)^{2} ]^{k-1}$, or to 
 $[\big(n^*_W/(1- n^*_W)\big) (W/\sigma)^{2} ]^{k-1}$  in the $\sigma_W$-expansion.   This
property  is important, since it implies that already our first-order result for $k=2$ from the $\sigma_W$-expansion  can describe the thermodynamic behavior even close to the jamming-transition density,  $n^*_\text{cp}(W) = \sigma/\sigma_W$, provided that the pore width is small enough such that the effective smallness parameter fulfills $\epsilon_W = [n^*_W/(1- n^*_W)]  (W/\sigma)^2  \ll 1$. Note that $\epsilon_W$ becomes the effective smallness parameter only close to $n^*_\text{cp}(W)$. Otherwise, it is the original parameter, $(W/\sigma)^{2}$.
  In order to illustrate the applicability of our cluster expansion, we have calculated analytically  the parallel and perpendicular pressure and the surface tension up to order $O(W^2)$.

Our work provides analytical results for the fundamental thermodynamic quantities that can be tested in experiments and computer simulations. These exact results should also serve as a benchmark for approximate theories or empirical approaches. As a demonstration, in Subsection~\ref{Sec:Comparison}, our leading-order result for the  parallel pressure is compared with both virial expansions \cite{Mon:PRE_97:2018} and with results from Monte-Carlo simulations \cite{ Kofke:JCP_98:1993,Mon:PRE_97:2018}. 

A comparison of the dimensionless pressure $p^*_\parallel$ [cf. Eq.~\eqref{eq:p-parallel}] with the  virial density series, Eq.~\eqref{eq:n-series},  yields \textit{all}  virial coefficients $B_l(W)$  up to order $O(W^2)$. Accordingly, already our  first-order result  for $p^*_\parallel$ corresponds to the sum of  the infinite number of virial terms  $B_{l+1}(W) (n^*)^l$(up to order $W^2 $). This underlines the strength of the present approach using the Tonks gas as an ``unperturbed'' fluid and not  the ideal gas, as in the standard virial expansion. Furthermore, for densities sufficiently below the close-packing density, $n^*_\text{cp}(W)$,  Fig.~\ref{fig:comparison_Mon}   shows that 
the first-order result \eqref{eq:p-parallel} is rather close to the virial pressure series, but deviates from the truncated virial density series. However, approaching $n^*_\text{cp}(W)$, $p^*_\parallel$ from the truncated virial pressure series saturates and fails to diverge at  $n^*_\text{cp}(W)$ in striking contrast to $p^*_\parallel$ from the $\sigma_W$-expansion \eqref{eq:p-parallel_W} (see Fig.~\ref{fig:comparison_Mon}). Since also $p^*_\parallel$ from the  $\sigma$-expansion fails to diverge (see Fig.~\ref{fig:comparison_Mon}), this  proves the quality and physical relevance of the $\sigma_W$-expansion in comparison with the  $\sigma$-expansion and both truncated virial series. 

Even up to values such as $W/\sigma = 0.8$, for which the smallness parameter is  not much smaller than unity,  the  comparison of the leading-order result of the $\sigma$- and $\sigma_W$-expansion [Eqs.~\eqref{eq:p-parallel}, ~\eqref{eq:p-parallel_W}] with the Monte-Carlo data \cite{ Kofke:JCP_98:1993,Mon:PRE_97:2018} 
 in Fig.~\ref{fig:comparison_Kofke} and Fig.~\ref{fig:comparison_Mon}   shows  a more or less good agreement for densities up to $n^*=0.6$.
 Due to the exponential increase of the pressure for $W/\sigma = 0.80$~\cite{Mon:PRE_97:2018}, there are not enough Monte-Carlo points in  Fig.~\ref{fig:comparison_Mon} to allow for a detailed comparison.  Therefore, further Monte-Carlo simulations would be desirable where $n^*$ and $W/\sigma$ are changed systematically.  In addition, such simulations also offer the possibility to calculate   the density profile and its comparison with our analytical result \eqref{eq:profile-q1d}. It  would also be interesting to elaborate the range of validity of our expansions, in particular for densities $n^*$ close to the jamming transition, choosing $W/\sigma$ small enough such that $\epsilon_W \ll  1$. 

 Our approach also constitutes a suitable starting point to calculate structural quantities for fluids in confined pores, similar to the case of a 2D slit pore~\cite{Lang:JCP_140:2014}. For example, the pair-distribution function for the parallel degrees of freedom can be evaluated in computer simulations and compared to integral-equation theories relying on the effective potentials derived in this work. In contrast, similar to the Asakura-Oosawa model,  the mapping to effective potentials is not designed to study general relaxation phenomena or dynamical properties, such as single-file diffusion~\cite{Wei:Science_287:2000}, although it may yield physically interesting results.

\begin{acknowledgments}
 We would like to thank  D.A.\@ Kofke, H.\@ L{\"o}wen   and   K.K.\@ Mon for  helpful discussions. This research was funded  in part  by the Austrian Science Fund (FWF) 10.55776/I5257. For open access purposes, the author has applied a CC BY public copyright license to any author accepted manuscript version arising from this submission.

\end{acknowledgments}

\section*{Author declarations}
\subsection*{Conflict of Interest}
 The authors have no conflicts to disclose.
\appendix

\section{Free energy for point particles}\label{Sec:Appendix_A}

In this appendix we supplement our analysis of the main text for particles interacting via a hard-core exclusion potential and hard walls by considering smooth interparticle interactions and smooth walls. The main strategy is now to develop a cumulant expansion~\cite{Hansen:Theory_of_Simple_Liquids:2023} rather than relying on a cluster expansion. 

The particles interact with the walls via a one-particle potential $\sum_i U(i_\perp, W)$ 
where $U(i_\perp, W) \equiv U(z_i, W)$ is a family of   smooth wall potentials depending only on the magnitude of the  transverse coordinate $z_i = |\mathbf{z}_i|$,  such that typical configurations fulfill $z_i< W/2$ with the effective pore width $W$.   Thus, configurations not fulfilling this constraint contribute with negligible weight in the thermodynamic limit and can be ignored in the subsequent analysis.
 Upon decreasing $W\to 0$ the wall potential $U( \cdot, W)$ becomes more and more confining.  
We introduce the transverse volume of the pore $V_\perp = \Omega_{\perp} (W/2)^{d_\perp}/d_\perp$, where $\Omega_\perp$ denotes the surface of the $d_\perp$-dimensional unit sphere, and consider $V_\perp$ as mechanical control parameter such that the free energy $F = F(T, V_\parallel, V_\perp, N)$ becomes a function of the natural variables $T, V_\parallel, V_\perp, N$.  The  analysis outlined below will allow us to provide a precise definition of the effective pore width in terms of the mean-squared excursions of a single particle.   

The total interaction potential  is then  provided by
\begin{align}\label{eq:total_potential_smooth}
V_{\text{tot}}(1\ldots N) = \sum_{i} U(i_\perp, W) + \sum_{i< j } v(ij) ,
\end{align} 
where 
the smooth two-body interaction
$v(ij) \equiv v(|\mathbf{r}_i-\mathbf{r}_j|)$ is  assumed to only depend  on the relative distance $|\mathbf{r}_i-\mathbf{r}_j|$. We focus on the regime where the pore width is small compared to the range of of the two-body potential.  

Then (for a typical configuration) the transverse distance of a pair fulfills $ z_{ij} := |\mathbf{z}_i-\mathbf{z}_j| < W$ and will be much smaller than the lateral distance $x_{ij} := |\mathbf{x}_i-\mathbf{x}_j|$. Therefore, one may rely on  the Taylor expansion,
\begin{equation}\label{eq:distance}
|{\mathbf{r}}_i - {\mathbf{r}}_j|  = x_{ij} 
 \left[1 + \frac{ z_{ij}^2}{2 x_{ij}^2}  - \frac{z_{ij}^4}{8  x_{ij}^4 }     +  O(W^6) \right]  .
\end{equation}
This observation allows also for expanding  the interparticle potential,
\begin{equation}\label{eq:pot-expan-1}
v(ij ) = \sum_{l=0}^{\infty}  v_l(ij) , 
\end{equation}
where  $v_l = O(W^{2l})$. The first few terms are 
\begin{align}\label{eq:pot-expan-2}
v_0(ij )&=  v(x_{ij}) ,  \nonumber \\
v_1(ij) &=  v'(x_{ij} )  \frac{z_{ij}^2}{2 x_{ij} } , \nonumber  \\
v_2(ij)&=   \left[x_{ij} v''(x_{ij} ) - v'(x_{ij})    \right] \frac{ z_{ij}^4}{ 8 x_{ij} ^3} .
\end{align}
Then, the  total  potential  can be rewritten as 
\begin{align}\label{eq:pot-expan-3}
V_{\text{tot}}(1\ldots N)  =&  \sum_i U(i_\perp) + V_0(1_\parallel\ldots N_\parallel) 
\nonumber \\ 
& + 
 \sum_{l=1}^\infty V_l(1\ldots N) , 
\end{align} 
where 
\begin{align}
V_0(1_\parallel\ldots N_\parallel) =  \sum_{i< j}v (|\mathbf{x}_i-\mathbf{x}_j|) ,
\end{align}
involves only the lateral degrees of freedom. The contribution of the first and second term of Eq.~\eqref{eq:pot-expan-3} will serve as our reference case where transverse and lateral d.o.f.\@ are independent. 
The coupling is introduced for $l\geq 1$ via  
\begin{align}\label{eq:coupling_potentials}
V_l(1\ldots N) = \sum_{i< j} v_l(ij) ,
\end{align}
where the weight of  the  interaction potentials  $V_l = O(W^{2l})$ becomes small as the pore size is decreased $W\to 0$.   

The $N$-particle density is defined as 
\begin{align}\label{eq:N-particle-density}
\rho^{(N)}(1\ldots N) = \exp[- \beta V_{\text{tot}}(1\ldots N) ] / Z^{\text{ex}} ,
\end{align}
where the excess partition function $Z^{\text{ex}} = Z^{\text{ex}}(T, V_\parallel, V_\perp, N)$ is provided by 
\begin{align}\label{eq:excess_partition}
Z^{\text{ex}}  := \int \Big[  \prod_{j=1}^N  \frac{\diff j}{V} \Big] \exp[- \beta V_{\text{tot}}(1\ldots N) ] .
\end{align}
Here, $V := V_\parallel V_\perp$ is the effective volume of the system. The excess free energy is obtained as $F^{\text{ex}} = F^{\text{ex}}(T,V_\parallel, V_\perp, N)= - k_B T \ln Z^{\text{ex}}$.  The total Helmholtz free energy is then $F = F^{\text{id}} + F^{\text{ex}}$ with the ideal gas contribution provided in Eq.~\eqref{eq:id-free-energy}. 

Structural averages with respect to the $N$-particle density, Eq.~\eqref{eq:N-particle-density}, are 
indicated by $\langle \ldots \rangle$. 
We introduce averages over the transverse degrees of freedom with respect to the wall potential by 
\begin{align}
\langle (\ldots) \rangle_\perp = \frac{1}{Z_\perp^{\text{ex}}} \int \Big[ \prod_{j=1}^N  \frac{\diff j_\perp}{V_\perp} 
\exp[ -\beta U(j_\perp)] \Big] \left( \ldots \right)    ,  
\end{align}
with the corresponding excess partition function 
\begin{align}
Z_\perp^{\text{ex}} = Z_\perp^{\text{ex}}(T,V_\perp, N) := \left[ \int \frac{\diff \mathbf{z}}{V_\perp} \exp[- \beta U(|\mathbf{z}|) ] \right]^N. 
\end{align}
Averages over the lateral degrees of freedom with respect to the reference potential $V_0(1\ldots N)$ are defined as in the main text 
[Eq.~\eqref{eq:definition_parallel_average}]  
\begin{align}
\langle (\ldots) \rangle_\parallel = \frac{1}{Z_\parallel^{\text{ex}}} \int \Big[ \prod_{j=1}^N  
\frac{\diff j_\parallel}{V_\parallel} \Big]  
\exp[ -\beta V_0(1_\parallel\ldots N_\parallel ) ] \left( \ldots \right)    ,  
\end{align}
with the associated partition function  
\begin{align}
Z_\parallel^{\text{ex}} = Z_\parallel^{\text{ex}}(T,V_\parallel, N) :=  \int \Big[ \prod_{j=1}^N  \frac{\diff j_\parallel}{V_\parallel} \Big] 
\exp[ -\beta V_0(1_\parallel\ldots N_\parallel ) ] .
\end{align}

Substituting Eq.~\eqref{eq:pot-expan-3} for the total potential in Eq.~\eqref{eq:excess_partition}  allows rewriting  the excess partition function  as 
\begin{align}
Z^{\text{ex}} &= Z^{\text{ex}}_\parallel Z^{\text{ex}}_\perp \langle \langle \exp[ -\beta \sum_{l=1}^\infty V_l(1 \ldots N) ] \rangle_\perp \rangle_\parallel .
\end{align}
Since $V_l = O(W^{2l})$, we can arrange the average of the exponential in terms of a cumulant expansion~\cite{Hansen:Theory_of_Simple_Liquids:2023} 
\begin{align}
\langle \langle \exp[ -\beta \sum_{l=1}^\infty V_l(1 \ldots N) ] \rangle_\perp \rangle_\parallel  =& \exp(- \beta \sum_{k=1}^\infty F_k) ,
\end{align}
where we follow the notation of the main text such that  $F_k = F_k(T,V_\parallel, V_\perp, N) = O(W^{2k})$.   
Explicitly, the first two  leading corrections are formally provided by 
\begin{eqnarray}\label{eq:cumulants}
- \beta F_1 &=&   -\beta \langle \langle  V_1 \rangle_{\perp} \rangle_{\parallel}   \ , \nonumber\\
-\beta F_{2}&=&  \beta^2/2 \big[\langle \langle  (V_1)^2 \rangle_{\perp} \rangle_{\parallel} 
- (\langle \langle  V_1 \rangle_{\perp} \rangle_{\parallel})^2 \big]   - \beta  \langle \langle  V_2 \rangle_{\perp} \rangle_{\parallel} . \nonumber \\
\end{eqnarray}
The total Helmholtz free energy is then obtained as 
\begin{align}
F = F^{\text{id}} + F^{\text{ex}}_0 + \sum_{k=1}^\infty F_k ,
\end{align}
where $F^{\text{ex}}_0 = F^{\text{ex}}_0(T, V_\parallel, V_\perp, N) :=  - k_B T \ln Z_\parallel^{\text{ex}} - k_B T\ln Z_\perp^{\text{ex}}$ is the excess free energy of the uncoupled reference system. 

In the following, we work out explicitly the leading correction $F_1 = O(W^2)$ of Eq.~\eqref{eq:cumulants}. 
The average over the lateral degrees of freedom can be performed relying on the pair-distribution function of the lateral reference system, Eq.~\eqref{eq:pair-distribution}. Following the same steps as in Appendix~\ref{Sec:Appendix_A}, we find the analog 
of Eq.~\eqref{eq:average-Phi}
\begin{align}
\langle \langle V_1 \rangle_\perp \rangle_\parallel = \frac{n N}{2} \int d \mathbf{x}   \, g(x)  \langle z_{12}^2 \rangle_\perp \frac{v'(x)}{x} ,
\end{align}
where $x= |\mathbf{x}|$. Since $\langle z_{12}^2 \rangle_\perp$ does not depend on the lateral separation of a pair of particles it can pulled in front of the integral. Using spherical symmetry, the integral simplifies to 
\begin{align}\label{eq:average_V1}
\langle \langle V_1 \rangle_\perp \rangle_\parallel = \frac{n N}{2} \langle z_{12}^2   \rangle_\perp  \Omega_{\parallel}  \int_0^\infty d x\,   x^{d_\parallel-2} g(x)  v'(x)   . 
\end{align}
 The  term $\langle z_{12}^2 \rangle_\perp$ quantifies the mean-squared transverse distance of a pair of particles in an ideal gas confined by the wall potential. Similar to Eq.~\eqref{eq:z_average}, symmetry dictates $\langle z_{12}^2 \rangle = 2 \langle z_1^2 \rangle$. We now use the mean-squared excursion of a single particle to \emph{define} the effective pore $W$ width by imposing
\begin{align}\label{eq:mean-squared}
\langle z_1^2\rangle_\perp = \frac{d_\perp}{d_\perp +2 } (W/2)^2 ,
\end{align}
such that it matches the limiting case of hard walls, Eq.~\eqref{eq:z_average}.  Collecting results of Eqs.~\eqref{eq:cumulants}, \eqref{eq:average_V1}, and \eqref{eq:mean-squared}, we find the exact expression for the leading correction
\begin{align}\label{eq:F-final3}
F_1& =     N  n \Omega_{\parallel} \frac{d_\perp}{8 (d_\perp+2)}    W^2 \int^{\infty}_{0} \diff x \,  x^{d_\parallel -2}  g(x) v'(x)  ,
\end{align}
which is the result presented in Eq.~\eqref{eq:finalF1-point}.

\section{Effective two-body potential}\label{Sec:Appendix_B}

In this appendix, we derive explicit expressions for the effective two-body potential of a confined fluid for the cases of one or two confining directions $d_\perp =1, 2$ and arbitrary number of unconfined directions $d_\parallel$.  

A formal expression for the effective two-body potential  was derived in Eq.~\eqref{eq:effpot6},
\begin{align}\label{eq:2-eff}
 v^{(2)}_{\text{eff}}(x)
 &= -k_B T \ln \left[1+  \left\langle \left[1 + f(x,z_{12}) \right] \right\rangle_\perp \right],
\end{align}
with the cluster function provided in Eq.~\eqref{eq:cluster1}. 
Since the cluster function vanishes  for $x < \sigma_W$ or   $x > \sigma$, the effective two-body potential is zero in these regimes. The remaining discussion is for the interval $\sigma_W \leq x \leq \sigma$.  There the cluster function simplifies to  
\begin{align}
f(x,z_{12}) = \Theta(z_{12}^2 - \sigma(x)^2) -1 ,
\end{align}
where we abbreviated $\sigma(x) := \sqrt{\sigma^2 -x^2} $.  Then, 
\begin{align}
\lefteqn{ 1 + \langle f(x, z_{12}) \rangle_\perp = } \nonumber \\
=& \int_{|\mathbf{z}_1 | < W/2} \frac{\diff \mathbf{z}_1}{V_\perp} \int_{|\mathbf{z}_2 | < W/2} \frac{\diff \mathbf{z}_2}{V_\perp} \Theta(z_{12}^2 - \sigma(x)^2) , 
\end{align}
with the  perpendicular distance $z_{12}= |\mathbf{z}_1-\mathbf{z}_2|$. Changing variables to the relative coordinate $\mathbf{z}_{12} = \mathbf{z}_1-\mathbf{z}_2$ yields
\begin{align}\label{eq:average_cluster}
\lefteqn{1 + \langle f(x, z_{12}) \rangle_\perp =} \nonumber \\
&= \frac{1}{V_\perp^2} \int_{|\mathbf{z}_{12} | < W} \diff \mathbf{z}_{12}   \Theta(z_{12}^2 - \sigma(x)^2) \int_{\mathcal{D}}  \diff \mathbf{z}_2. 
\end{align}
Here, the domain of the inner integral is 
\begin{align}
\mathcal{D} = \{ \mathbf{z}_2 \in \mathbf{R}^{d_\perp}: |\mathbf{z}_2| < W/2   \land |\mathbf{z}_{12} + \mathbf{z}_2 | < W/2
\} ,
\end{align}
and corresponds geometrically to the intersection of two $d_\perp$-dimensional spheres of radii $W/2$ separated by a distance $z_{12}$. We denote the volume of this domain  by $|\mathcal{D}|$. Here, we restrict the discussion to the most relevant cases of $d_\perp=1$ and $d_\perp =2$. 

\subsection{One confining direction} 

For $d_\perp=1$, the intersection is merely an interval of length $|\mathcal{D} |= W- z_{12}$. The remaining integral in Eq.~\eqref{eq:average_cluster} is elementary, and with $V_\perp = W$, we find
\begin{align}\label{av_cluster_app}
1 + \langle f(x, z_{12}) \rangle_\perp =& \frac{2}{W^2} \int_{\sigma(x)}^{W} \diff z_{12} ( W- z_{12} )  \nonumber \\
=& \left[1- \sigma(x)/W \right]^2    .
\end{align}
The effective potential then follows from Eq.~\eqref{eq:2-eff} to 
\begin{align}\label{eq:v2_Appendix}
v^{(2)}_{\text{eff}}(x) = -2 k_B T \ln[ 1 - \sqrt{(\sigma^2 -x^2) /W^2 } ] ,
\end{align} 
which is Eq.~\eqref{eq:effpot7} 
of the main text. 

\subsection{Two confining directions} 
For $d_\perp=2$, the intersection corresponds to a symmetric lens and its area can be expressed via the difference of a sector and a triangle. Elementary geometry yields
\begin{align}
|\mathcal{D}| =   \frac{W^2}{2} ( \alpha -  \cos \alpha \sin \alpha )  ,
\end{align}
with $\cos \alpha = z_{12}/W$. 
The remaining integral in Eq.~\eqref{eq:average_cluster} is conveniently evaluated in polar coordinates
\begin{align}
1 + \langle f(x, z_{12}) \rangle_\perp =& \frac{1}{V_\perp^2}  \int_{\sigma(x)}^W 2\pi z_{12}\diff z_{12}\, |\mathcal{D}| .
\end{align}
Upon change of variables $ z_{12} = W \cos\alpha$ and observing $V_\perp = \pi (W/2)^2$, the integral reads
\begin{align}
1 + \langle f(x, z_{12}) \rangle_\perp =&  \frac{32}{\pi W^2}   \int_{0}^{\alpha(x)} \cos\alpha  \sin \alpha \, \diff \alpha \, |\mathcal{D}| ,  
\end{align}
with $\cos\alpha(x) = \sigma(x) /W$. The last integral is elementary and yields 
\begin{align}\label{eq:cluster_function_app}
1 + \langle f(x, z_{12}) \rangle_\perp =&  \frac{2}{\pi} \frac{\sigma(x)}{W} \left( 1+ 2 \Big( \frac{\sigma(x)}{W}\Big)^2 \right) \sqrt{1- (\sigma(x)/W)^2} \nonumber \\
&  + \frac{2}{\pi} \left(1- 4 \Big(\frac{\sigma(x)}{W} \Big)^2 \right) \arccos\big( \sigma(x)/W \big) .  
\end{align}
 The effective two-body potential $v_{\text{eff}}^{(2)}(x)$ then follows upon substituting this result into Eq.~\eqref{eq:2-eff}.  It is presented in  Fig.~\ref{fig:soft_2_body_potential}.

\section{Effective three-body potential}\label{Sec:Appendix_C}

 In this appendix, we calculate the effective three-body potential for the case of a q1D fluid for a single confining direction $d_\perp =1$. 
 
 The effective three-body potential was given by Eq.~\eqref{eq:eff-3-particle} in the main text, and can be rewritten equivalently as
\begin{align}
 \lefteqn{ \exp  \left[ -\beta v^{(3)}_{\text{eff}}(1_\parallel 2_\parallel 3_\parallel) 
\right]  = } \nonumber \\
&=   \frac{ \left\langle [ 1+ f(12) ] [1+ f(23) ] [1+ f(13)] \right\rangle_\perp }{  
[ 1+\langle f(12) \rangle_\perp] [1+\langle f(23) \rangle_\perp ]
 [1+ \langle f(13) \rangle_\perp ] } \ ,
\end{align}
where Eq.~\eqref{eq:effpot6}
 has been used.
By translational invariance it only depends   on the relative parallel positions of the particles, say $\mathbf{x}_{12} = \mathbf{x}_1-\mathbf{x}_2, \mathbf{x}_{23} = \mathbf{x}_{2}-\mathbf{x}_3$. The three-body potential 
 generally requires as new input the calculation of the average of the product of three cluster functions in the numerator. The inclusion of the hard-sphere repulsion with exclusion distance $\sigma_W$ in the perpendicular average $\langle \ldots \rangle_\perp$ suppresses all configurations where  the parallel distance $x_{ij} = |\mathbf{x}_i -\mathbf{x}_j|$  of any of the pairs $(ij)$ is smaller than $\sigma_W$.  Furthermore, the cluster function $f(ij)$ vanishes for parallel distance $x_{ij} > \sigma$. For $\sigma_W <  x_{ij} < \sigma$, the cluster function $f(ij)$ is nonvanishing and we shall say that 
the particle pair $(i j)$ is connected by  a 'bond'. Generally, the effective potential is non-vanishing only if each particle in the cluster is connected by a bond to some other particle in the cluster. For the case of three particles, upon permutation of the labels, we may assume that there are bonds $(12), (23)$. There are two cases to be considered, either the bond $(13)$ is present or not. This is illustrated in  Fig.~\ref{fig:cluster_expansion}(c)  and Fig.~\ref{fig:cluster_expansion-1d}(c). The three-cluster on the left of  Fig.~\ref{fig:cluster_expansion}(c) and Fig.~\ref{fig:cluster_expansion-1d}(c)  consists of three bonds, whereas the second one from the left displays only two bonds.

An elementary geometrical consideration reveals that the case of three simultaneous  bonds $(12), (23), (13)$,  can only occur  for wide enough pores $W/\sigma \geq \sqrt{3}/2 \approx 0.86$.  Since we are mainly  interested in the case of narrow confinement, we consider only the case where the bond $(13)$ is not present, i.e., the distance fulfills $x_{13}> \sigma$ such that $f(13)=0$. 
Furthermore, we restrict the discussion to  $d_{\perp}=1$; the subsequent calculation should be considered as proof or principle. In this case, the three-body potential reduces to 
\begin{align}\label{eq:3-pot-reduced}
  \exp  \left[ -\beta v^{(3)}_{\text{eff}}(x_{12}, x_{23}) 
\right]  =   \frac{ \left\langle [ 1+ f(12) ] [1+ f(23) ]\right\rangle_\perp }{  
[ 1+\langle f(12) \rangle_\perp] [1+\langle f(23) \rangle_\perp ] } .
\end{align}
Our notation already made use of the fact that in the one-dimension case, we can sort the particles $x_1 < x_2 < x_3$ such that $x_{12} = x_2-x_1 > 0, x_{23} = x_3-x_2$ are positive. The numerator in Eq.~\eqref{eq:3-pot-reduced} then is provided by the integral 
\begin{align}\label{eq:3-integral}
\lefteqn{I:= \left\langle [ 1+ f(12) ] [1+ f(23) ]\right\rangle_\perp = } \nonumber \\
= & \int_{-W/2}^{W/2}\! \frac{\diff z_1}{W} \! \int_{-W/2}^{W/2}\! \frac{\diff z_2}{W}  \! \int_{-W/2}^{W/2}\! \frac{\diff z_3}{W} 
\Theta( |z_1-z_2|^2 - \sigma(x_{12})^2 )  \nonumber \\
&\times \Theta( |z_2-z_3|^2 - \sigma(x_{23})^2 )  .
\end{align}
We now switch to relative coordinates $\zeta_{12} = z_1- z_2 , \zeta_{23} = z_2-z_3$ such that the integral simplifies to 
\begin{align} 
I = & \int_{-W}^{W}\! \frac{\diff \zeta_{12}}{W}  \Theta( \zeta_{12}^2 - \sigma(x_{12})^2 ) \int_{-W}^{W}\! \frac{\diff \zeta_{23}}{W} 
\Theta( \zeta_{23}^2 - \sigma( x_{23})^2 ) \nonumber \\
& \times  \int_{\mathcal{D}}\frac{\diff z_2}{W}   ,
\end{align}
where the integration domain $\mathcal{D}$ is the intersection of three intervals
\begin{align}
\mathcal{D} =&  [-W/2, W/2] \cap [-W/2 - \zeta_{12}, W/2-\zeta_{12} ]  \nonumber \\
& \cap [ -W/2 + \zeta_{23} , W/2 + \zeta_{23} ] .
\end{align}
We assume that $\zeta_{12}> 0$, the case $\zeta_{12}<0$ is obtained by reflecting all $z$-coordinates. For $\zeta_{23} >0$ the centers of the latter two intervals are on opposite sides of the origin and the intersection of the three intervals is $[\zeta_{23}-W/2, - \zeta_{12} + W/2]$ provided that $ -\zeta_{12}+ W/2 \geq \zeta_{23}-W/2$ and empty otherwise. In the other case $\zeta_{23}< 0$ both centers of the latter two intervals are to the left to the origin and the intersection of the three intervals is $[-W/2, \text{min}( \zeta_{23} + W/2, -\zeta_{12} + W/2) ]$. 
Collecting results, the length of the interval is therefore given by
\begin{align}
|\mathcal{D}| = \begin{cases} 
\text{max}( W - \zeta_{12} - \zeta_{23} , 0 )  & \text{for } \quad \zeta_{23} > 0,  \\
W - \text{max}(-\zeta_{23}, \zeta_{12} )   & \text{for } \quad \zeta_{23} < 0.
\end{cases}
\end{align}
Then the integral in Eq.~\eqref{eq:3-integral} becomes
\begin{align}\label{eq:Isum}
I :=& I_+ + I_- \nonumber \\
= &  \frac{2}{W^3} \int_{\sigma(x_{12})} ^W \diff\zeta_{12} \left(  \int_{\sigma(x_{23})}^{W} + \int_{-W}^{-\sigma(x_{23}) } \right) \diff \zeta_{23}\,  |\mathcal{D}| .
\end{align}

The integral over positive $\zeta_{23}$ is straightforward 
\begin{align}\label{eq:Iplus}
I_+ = \text{max}\left( \frac{1}{3} [1- \sigma(x_{12})/W-\sigma(x_{23})/W]^3, 0 \right)  ,
\end{align}
and involves an additional condition on the parallel distances. The condition could have been read off directly from the definition in Eq.~\eqref{eq:3-integral} since the integrand is nonvanishing only for $\sigma(x_{12}) + \sigma(x_{23}) \leq \zeta_{12} + \zeta_{23} = z_1-z_3  \leq W$.   

For the integral $I_-$, we need to split the inner integral into parts where $\zeta_{12} \gtrless \sigma(x_{23})$, and for the outer integral, we have to consider the two cases $\sigma(x_{12}) \gtrless \sigma(x_{13})$.  The final result for $\sigma(x_{12}) > \sigma(x_{23})$ is 
\begin{align}\label{eq:Iminus}
I_- = \frac{1}{3W^3 } [ W- \sigma(x_{12}) ]^2  [ \sigma(x_{12}) -3 \sigma(x_{23}) + 2 W ] .
\end{align} 
 For the other case $\sigma(x_{12}) < \sigma(x_{23})$, the same expression holds upon interchanging $\sigma(x_{12}) \leftrightarrow \sigma(x_{23}) $. Substituting the results of  Eqs.~\eqref{eq:Iminus}, \eqref{eq:Iplus}, \eqref{eq:Isum}, \eqref{eq:3-integral} as well the expression for the effective two-body potential, Eq.~\eqref{eq:effpot7}, into Eq.~\eqref{eq:3-pot-reduced},
 one obtains the effective three-body potential $v^{(3)}_{\text{eff}}(x_{12}, x_{23} )$ 
 for the case of two bonds. Although particles 1 and 3 do not interact with each other directly, the effective three-body potential is non-zero, highlighting that the interaction energy of the three-cluster is not simply the sum of the effective two-body potentials for the bonds. This is rather different from the standard virial expansion where a three-cluster consisting only of two bonds would be considered as 'reducible' in the sense that it is essentially a composition of a pair of of two-clusters.
The three-body potential for $ W =0.75 \sigma < W_1$, i.e., where only two bonds can exist, is displayed in   Fig.~\ref{fig:soft_3_body_potential}.

\section{Averages of the cluster functions}\label{Sec:Appendix_D}
In this appendix we calculate the average of the cluster function $\sum_{i<j}\langle \langle f(ij)\rangle_\perp\rangle_\parallel$  to leading order $O(W^2)$ in the pore width. Consider the observable in parallel configuration space
\begin{align}\label{eq:D1}
\Phi = \Phi(1_\parallel\ldots N_\parallel) := \sum_{i<j} \langle f(ij) \rangle_\perp ,
\end{align}
which is a two-body operator, i.e., it is a sum over all pairs of particles. The goal is therefore to calculate $\langle \Phi \rangle_\parallel$.  
 Progress is made by defining the pair-distribution function~\cite{Hansen:Theory_of_Simple_Liquids:2023} of the $d_\parallel$-dimensional  reference system
\begin{align}\label{eq:pair-distribution}
g(x) := \frac{1}{n N} \big\langle \sum_{i \neq j}  \delta(\mathbf{x} - (\mathbf{x}_i-\mathbf{x}_j)) \big\rangle_\parallel, 
\end{align}
where  $n=N/V_\parallel$ is the density of the reference fluid. The quantity $g(x)$ corresponds to the expected number of particles separated by a vector $\mathbf{x}$ from a selected particle relative to an ideal gas at the same density. 
Note that the dependence of $g(x)$ on the diameter $\sigma_W$ of the hard spheres of the reference fluid has been suppressed.
We have exploited already that the fluid is translationally  and rotationally invariant in the thermodynamic limit such that the pair-distribution function only depends on the magnitude $x= |\mathbf{x}|$. 
Then, the average is obtained as for any two-body operator 
\begin{align}
 \langle \Phi \rangle_\parallel &= \frac{n^2}{2} \int\!\diff \mathbf{x}_1\!\int\! \diff  \mathbf{x}_2 \,   g(x_{12}) \langle f(x_{12}, z_{12}) \rangle_\perp  ,
 \end{align}
 where $x_{12} = |\mathbf{x}_1-\mathbf{x}_2|$, $z_{12} = |\mathbf{z}_1-\mathbf{z}_2|$ are the distances of the pair in  the parallel and perpendicular direction. 
By translational invariance in the parallel direction, one can introduce the  relative coordinate $\mathbf{x} =  \mathbf{x}_1-\mathbf{x}_2$
 and evaluate the integral to 
\begin{align} \label{eq:average-Phi}
 \langle \Phi \rangle_\parallel &= \frac{n N}{2} \int \diff \mathbf{x}  \,  g(x ) \langle f( x, z_{12}) \rangle_\perp  ,
\end{align}
 The integrand only depends  on  $x=|\mathbf{x}|$ rather than $\mathbf{x}$ such that the integral simplifies in spherical coordinates. 
Since the support of $\langle f(x,z_{12}) \rangle_\perp$ is restricted to the narrow shell{ $\sigma_W \leq x \leq \sigma$,  we can approximate the regularly varying terms $g(x)$ and $x^{d_\parallel-2}$ (part of the volume element) by its contact value  $g_+ = g_+(n^*) := g(x\downarrow \sigma) $ and $\sigma^{d_\parallel-2}$, respectively,  as $W\to 0$ in the integrand. 
The limit $W \to 0$  implies that $\sigma_W \to \sigma$; therefore, $g(x)$  becomes the pair-distribution function
of a fluid of hard spheres of diameter $\sigma$ and  $g_+(n^*)$ is the corresponding value at contact $x= \sigma$. 
We emphasize its dependence on the reduced density $n^* = N \sigma^{d_\parallel}/ V_\parallel$ to avoid confusion.  We arrive to leading order at
\begin{align} 
 \langle \Phi \rangle_\parallel &= \frac{n N}{2} g_+(n^*)  \sigma^{d_\parallel-2} \Omega_\parallel \int_{\sigma_W}^\sigma    
\langle f(x,z_{12}) \rangle_\perp  x \diff x .
\end{align}
The remaining regular factor $x$ of the volume element is kept  in the integrand to simplify the subsequent manipulations. Here, $\Omega_{\parallel}$ denotes the  surface of the $d_\parallel$-dimensional unit sphere.  
In the range of the integration the cluster function simplifies to $f(x,z_{12}) = \Theta(x^2 + z_{12}^2 - \sigma^2) -1$. Interchanging the perpendicular average with the integral yields to leading order
\begin{align}\label{eq:Phi_av}
 \langle \Phi \rangle_\parallel &= \frac{n N}{2} g_+(n^*) \sigma^{d_\parallel-2} \Omega_{\parallel} \  \cdot \nonumber\\
 & \cdot \left\langle \int_{\sigma_W}^\sigma \left[\Theta(x^2 +z_{12}^2 - \sigma^2) -1 \right]x \diff x \right\rangle_\perp  \nonumber \\
 &= \frac{n N}{4} g_+(n^*) \sigma^{d_\parallel-2} \Omega_{\parallel} \left( \langle   z_{12}^2 \rangle_\perp- W^2 \right) .
\end{align}
The average over the perpendicular quadratic excursion  is readily computed, observing that in $\langle z_{12}^2 \rangle_\perp = \langle z_1^2 -2 \mathbf{z}_1 \cdot \mathbf{z}_2 + z_2^2\rangle_\perp$ the mixed term does not contribute. Using spherical coordinates, we find
\begin{align}\label{eq:z_average}
\langle |\mathbf{z}_1 - \mathbf{z}_2| ^2 \rangle_\perp = 2 \langle z_1^2 \rangle_\perp  = \frac{2 d_\perp}{d_\perp+2} (W/2)^2 . 
\end{align}
Collecting results, we find
 \begin{align}
 \langle \Phi \rangle_\parallel & = -N \frac{n^* }{8} g_+(n^*)  \Omega_{\parallel} \frac{d_\perp+ 4}{d_\perp+2} \left( \frac{W}{\sigma}\right)^2 + O(W^4) ,
 \end{align} 
 which yields Eq.~\eqref{eq:clusterF-final}. 
 
 The previous calculation is identical for the calculation of $\langle f(12) \rangle_\perp' \rangle_\parallel$ required for calculating the density profile, Eq.~\eqref{eq:profile2}, where the prime indicates that no average is performed over $\mathbf{z}_1$.  By permutation symmetry we find from Eq.~\eqref{eq:Phi_av} to leading order
 \begin{align}
 \langle f(12) \rangle_\perp' \rangle_\parallel =& \frac{n}{2 (N-1)} g_+(n^*) \sigma^{d_\parallel-2} \Omega_{\parallel} \nonumber \\
& \times \int\frac{\diff \mathbf{z}_2}{V_\perp} \left[    z_1^2 -2 \mathbf{z}_1 \cdot \mathbf{z}_2 + z_2^2  - W^2 \right] .
 \end{align}
By symmetry, the scalar product $\mathbf{z}_1 \cdot \mathbf{z}_2$ averages out. Subtracting the averaged cluster function  
we arrive at
\begin{align}
\lefteqn{ \langle f(12) \rangle_\perp' \rangle_\parallel
-  \langle f(12) \rangle_\perp \rangle_\parallel } \nonumber \\
 =& \frac{n^*}{2 (N-1)} g_+(n^*)  \Omega_{\parallel} \sigma^{-2}  [ z_1^2 - \langle z_1^2 \rangle_\perp] + O(W^4) . 
\end{align} 
With Eq.~\eqref{eq:z_average}, this yields the density profile, Eq,~\eqref{eq:profile_final}, of the main text.

The results above correspond to the $\sigma$-expansion. Since the $\sigma_W$-expansion
even encompasses the divergence of, e.g., the parallel pressure at the physical correct closed-packing density, we 
will recalculate the average in Eq.~\eqref{eq:average-Phi} using a slightly different route. This will be done for  $d_\parallel = 1$, since we have used results from the $\sigma_W$-expansion for the q1D fluid, only. 
As a benefit, we also identify the effective smallness parameter in this expansion.
Using reflection symmetry, Eq.~\eqref{eq:average-Phi} becomes   
\begin{align} \label{eq:average-Phi-W-1}
 \langle \Phi \rangle_\parallel &=  n_W^* N \int_{1}^{\sigma/\sigma_W} \diff y  \,  g(y;\sigma_W ) \langle f(\sigma_W y, z_{12}) \rangle_\perp  ,
\end{align}
where $  g(y;\sigma_W ) $ indicates  the use of the Tonks gas with hard rods of length $\sigma_W$  as  reference fluid. Furthermore we introduced  $y=x/\sigma_W$ as the integration variable and abbreviated $n_W^* = n \sigma_W$.

For $W/\sigma < \sqrt{3}/2$  the calculation of the integral is rather simple since $1 \leq \sigma/\sigma_W < 2$.
In this domain the pair-distribution function is given by  \cite{Salsburg:JCP_21:1953} 
\begin{align}\label{eq:g-W}
g(y;\sigma_W ) =  \frac{1}{1 - n_W^*} \exp\Big[ \frac{ - n_W^* (y - 1)}{1-n_W^*} \Big] \  .
\end{align}
 Substitution into Eq.~\eqref{eq:average-Phi-W-1} allows calculating the integral.  Taking into account $f(\sigma_W y, z_{12}) = \Theta\big(\sigma_W^2 y^2 - \sigma(\mathbf{z}_{12})^2 \big) -1 $ and use of $\sigma(\mathbf{z}_{12}) = \sqrt{\sigma^2 - \mathbf{z}_{12}^2 }$,
one obtains
\begin{align} \label{eq:average-Phi-W-2}
 \langle \Phi \rangle_\parallel &= - N  
 \left[ 1 - \left\langle  \exp{\big[- \frac{n_W^*(\sigma (\mathbf{z}_{12})/\sigma_W - 1 ) }{1-n_W^*}   \big]} \right\rangle_\perp \right]  \ .
\end{align}
This result is still exact. 
Next, we expand with respect to  $W/\sigma$
\begin{align} \label{eq:average-Phi-W-3}
\sigma(\mathbf{z}_{12})/\sigma_W - 1 
=& \frac{(W/\sigma)^2}{2} \left[ 1 - \frac{\mathbf{z}_{12}^2}{W^2}   
  + O(W^2)    \right]  .
\end{align}
Introducing the effective smallness parameter $\epsilon_W := [n_W^*/(1-n_W^*)] (W/\sigma)^2$ one can expand   the exponential function in Eq.~\eqref{eq:average-Phi-W-2} with respect   to $\epsilon_W$ and finds
\begin{align} \label{eq:average-Phi-W-4}
& \langle \Phi \rangle_\parallel = -\frac{N}{2} \epsilon_W   \big[1 -  \langle \mathbf{z}_{12}^2\rangle_\perp / W^2 + O(W^2)\big]   +  O(\epsilon_W^2)   .
\end{align}
Finally, Eq.~\eqref{eq:z_average} implies  
\begin{align} \label{eq:average-Phi-W-5}
 \langle \Phi \rangle_\parallel &= - \frac{N}{4}  \frac{d_\perp + 4}{d_\perp+2} \epsilon_W  [ 1 + O(W^2)] +  O(\epsilon_W ^2)\ .
\end{align}
Substituting this result into Eq.~\eqref{eq:clusterF} yields  for $d_\perp=2$ the leading order result \eqref{eq:F_2-cluster_W}.


\section{Expansion of the excess free energy of the reference fluid  with respect to the pore width}
\label{Sec:Appendix_E}
We  expand the excess free energy of the reference fluid $F^{\text{ex}}_\parallel(T, V_\parallel, N; \sigma_W)$ consisting of (hyper-)spheres of  diameter $\sigma_W$ at number density $n = N/V_\parallel$ to leading nontrivial order in $W$. By dimensional analysis $F^{\text{ex}}_\parallel/ Nk_B T$ only depends  on the dimensionless variable $ N (\sigma_W)^{d_{\parallel}} / V_\parallel$.  Then, by the chain rule,
\begin{align}
\left( \frac{\partial F^{\text{ex}}_\parallel}{\partial \sigma_W} \right)_{T,V_\parallel, N} =&   - \frac{d_\parallel}{\sigma_W } V_\parallel \left(\frac{\partial F^{\text{ex}}_\parallel}{\partial V_\parallel}  \right)_{T,N, \sigma_W} \nonumber \\ =&  \frac{d_\parallel}{\sigma_W } 
V_\parallel \, p^{\text{ex}}_\parallel ,
\end{align}
where $p^{\text{ex}}_\parallel = p^{\text{ex}}_\parallel(T, V_\parallel, N; \sigma_W)$ denotes the excess pressure. By the help of  the virial theorem~\cite{Hansen:Theory_of_Simple_Liquids:2023}, the excess pressure evaluated for the reference fluid can be related directly to its contact value
\begin{align}\label{eq:virial_theorem}
p^{\text{ex}}_\parallel(T,V_\parallel, N; \sigma)  &= - \frac{n^2}{2 d_\parallel} \Omega_{\parallel} \int x u'(x) g(x) x^{d_{\parallel-1}} \diff x  \nonumber \\
& =  \frac{n^2}{2 d_\parallel} \Omega_{\parallel} k_B T \sigma^{d_\parallel} g_+(n^*)  .
\end{align}
Collecting results, we find to leading nontrivial order
\begin{align}
F_\parallel^{\text{ex}}(T,V_\parallel, N ; \sigma_W ) 
=& F_\parallel^{\text{ex}}(T,V_\parallel, N ; \sigma ) \nonumber \\
& - N k_B T \Omega_{\parallel} n^* g_+(n^* )  \frac{W^2}{4 \sigma^2} + O(W^4) .
\end{align}


%

\end{document}